\documentclass[runningheads]{llncs}

\usepackage[T1]{fontenc}
\usepackage{amssymb}
\usepackage[hidelinks]{hyperref}      
\usepackage{url}            
\usepackage{booktabs}       
\usepackage{amsfonts}       
\usepackage{nicefrac}       
\usepackage{microtype}      
\usepackage{lipsum}		    
\usepackage{doi}
\usepackage{subcaption}
\usepackage{colortbl}
\usepackage{balance}
\usepackage{xspace}
\usepackage{acronym}
\usepackage{adjustbox}
\usepackage{siunitx}
\usepackage{tabularx}
\usepackage{doi}
\usepackage[table,dvipsnames]{xcolor}
\usepackage{pifont}
\usepackage{multirow}
\usepackage[moderate]{savetrees}
\usepackage{enumitem}
\usepackage[strict]{changepage}
\usepackage{framed}

\definecolor{formalshade}{rgb}{0.85,1,0.85} 
\definecolor{darkblue}{rgb}{0.0,0.6,0.30}

\definecolor{formalshade}{rgb}{0.85,1,0.85}
\definecolor{darkblue}{rgb}{0.0,0.6,0.30}

\newboolean{showcomments}
\setboolean{showcomments}{true}
\ifthenelse{\boolean{showcomments}}
 { \newcommand{\mynote}[2]{
      \fbox{\bfseries\sffamily\scriptsize#1}
        {\small$\blacktriangleright$\textsf{\emph{#2}}$\blacktriangleleft$}}}
        { \newcommand{\mynote}[2]{}}

\usepackage{scalerel}
\usepackage{tikz}

\newcommand*\emptycirc[1][1ex]{\tikz\draw (0,0) circle (#1);} 

\newcommand*\fullcirc[1][1ex]{\tikz\fill (0,0) circle (#1);}

\begin{document}
% \title{DUMB and DUMBer:\\Is Adversarial Training Actually Worth It?}
\title{DUMB and DUMBer:\\Is Adversarial Training Worth It in the Real World?}
\titlerunning{DUMB and DUMBer}

\author{
Francesco Marchiori\inst{1}\orcidID{0000-0001-5282-0965} \and
Marco Alecci\inst{2}\orcidID{0000-0002-5963-4599} \and
Luca Pajola\inst{3}\orcidID{0000-0002-6749-6608} \and
Mauro Conti\inst{1,3}\orcidID{0000-0002-3612-1934}
}
\authorrunning{F. Marchiori et al.}
\institute{University of Padova, Padova, Italy \and
University Of Luxembourg, Luxembourg, Luxembourg \and
Spritz Matter Srl, Padova, Italy\\
\email{francesco.marchiori@math.unipd.it,~marco.alecci@uni.lu,
luca.pajola@spritzmatter.com,
mauro.conti@unipd.it}}

\maketitle  
\setcounter{footnote}{0}

\begin{abstract}
Adversarial examples are small and often imperceptible perturbations crafted to fool machine learning models.
These attacks seriously threaten the reliability of deep neural networks, especially in security-sensitive domains.
Evasion attacks, a form of adversarial attack where input is modified at test time to cause misclassification, are particularly insidious due to their transferability: adversarial examples crafted against one model often fool other models as well.
This property, known as adversarial transferability, complicates defense strategies since it enables black-box attacks to succeed without direct access to the victim model.
% While training models with adversarial examples remains one of the most common defense strategies, it is still unclear whether these techniques can consistently increase model robustness across varying scenarios, especially when tested under multiple transferability dimensions.
While adversarial training is one of the most widely adopted defense mechanisms, its effectiveness is typically evaluated on a narrow and homogeneous population of models.
This limitation hinders the generalizability of empirical findings and restricts practical adoption.

In this work, we introduce \textbf{DUMBer}, an attack framework built on the foundation of the DUMB (Dataset soUrces, Model architecture, and Balance) methodology, to systematically evaluate the resilience of adversarially trained models.
Our testbed spans multiple adversarial training techniques evaluated across three diverse computer vision tasks, using a heterogeneous population of uniquely trained models to reflect real-world deployment variability.
% Our testbed includes multiple popular adversarial training techniques and evaluates them across three diverse computer vision tasks.
Our experimental pipeline comprises over 130k evaluations spanning 13 state-of-the-art attack algorithms, allowing us to capture nuanced behaviors of adversarial training under varying threat models and dataset conditions.
Our findings offer practical, actionable insights for AI practitioners, identifying which defenses are most effective based on the model, dataset, and attacker setup.
% Our findings provide novel insights into the efficacy and limitations of adversarial training, revealing which practices are most effective depending on the type of adversary, model, and dataset setup and providing practical recommendations for building more robust machine learning systems.
\end{abstract}

\section{Introduction}
\label{sec:introduction}

Deep Neural Networks (DNNs) have achieved remarkable performance across a wide range of tasks, particularly in computer vision, natural language processing, and autonomous systems.
However, these models are known to be highly vulnerable to adversarial examples, i.e., carefully crafted inputs that cause the model to make incorrect predictions while appearing benign to human observers.
First identified in the context of image classification~\cite{goodfellow2014explaining}, adversarial attacks have since evolved into a rich area of research, encompassing various modalities and attack scenarios.
One prominent class of these threats is \textit{evasion attacks}, where the attacker modifies inputs at test time to evade detection or mislead the model.
Real-world manifestations of such attacks have been observed in malware detection systems~\cite{ling2023adversarial}, facial recognition spoofing~\cite{sharif2016accessorize}, and autonomous driving~\cite{eykholt2018robust}, raising significant concerns about the safety and reliability of AI systems deployed in adversarial environments.

A particularly troubling property of adversarial examples is their \textit{transferability}: adversarial inputs crafted for one model often succeed in misleading other models, even if they differ in architecture, training data, or optimization details~\cite{demontis2019adversarial,wang2021admix}.
This phenomenon enables gray-box and black-box attacks, where the attacker has limited or no access to the target model’s internals yet can still craft effective attacks using surrogate models.
As a result, transferability significantly undermines the security of models in deployed settings.
The DUMB framework~\cite{alecci2023dumb} was recently proposed to study how adversarial transferability varies as a function of three key dimensions: \underline{D}ataset so\underline{U}rces, \underline{M}odel architectures, and the \underline{B}alance of class distributions.
It provided a standardized way to evaluate how generalizable adversarial examples are across different training conditions, laying the foundation for more robust empirical evaluations of adversarial threats.

One widely studied defense mechanism against adversarial attacks is \textit{adversarial training}, where the model is trained on adversarial examples in addition to clean data~\cite{madry2017towards,andriushchenko2020understanding}.
This technique aims to increase the model’s robustness by explicitly teaching it to resist known types of perturbations.
While adversarial training has shown promise, especially in white-box settings, its effectiveness often comes with trade-offs: it can lead to significant reductions in accuracy on clean (non-adversarial) inputs~\cite{tsipras2018robustness}, increase training time, and sometimes fail to generalize beyond the specific attack types used during training.
Various improved techniques have been proposed~\cite{zhang2019theoretically,wang2019improving}, each attempting to balance robustness and standard accuracy.
However, while the individual performance of adversarially trained models has been extensively evaluated, the impact of adversarial transferability on these defenses remains underexplored.
Given the high computational demands of adversarial training and the practical challenges of deploying robust models in real-world settings, particularly under the threat of adversarial transferability, we are prompted to ask: \textit{is adversarial training truly effective in real-world scenarios?} And if so, \textit{which strategies offer the most resilience across diverse attack conditions?} This work seeks to answer these questions by systematically evaluating various adversarial training techniques under controlled yet transferable attack settings.

\paragraph{Contribution.}
This work presents \textbf{DUMBer}, an extension of the DUMB framework that evaluates the impact of adversarial training on transferability across models, datasets, and class balance conditions.
While DUMB analyzed the robustness of models to transferred evasion attacks in standard training settings, DUMBer investigates whether and how commonly used adversarial training techniques improve resilience under the same conditions.
We focus on the interaction between adversarial training and transferability, providing a systematic evaluation across a broad spectrum of configurations.
Our aim is to address a key limitation in the literature: adversarial training techniques are often evaluated on a narrow and homogeneous set of models, which is quite limiting given the empirical nature of these defenses.
In contrast, we adopt a ``\textbf{DUMB population}'' where every model is unique, better capturing the variability encountered in real-world deployments.
Our contributions can be summarized as follows.

\begin{itemize}
    \item We propose \textbf{DUMBer}, an attacker model and framework that extends DUMB by incorporating adversarial training into the evaluation of adversarial transferability.
    % Our testbed evaluates the effects of adversarial training across three axes: \textbf{D}ataset so\textbf{U}rces, \textbf{M}odel architectures, and class \textbf{B}alance.
    Across three axes (\underline{\textbf{D}}ataset so\underline{\textbf{U}}rces, \underline{\textbf{M}}odel architectures, and class \underline{\textbf{B}}alance), our testbed evaluates the \underline{\textbf{e}}vasion \underline{\textbf{r}}esilience of adversarially trained models.
    % \item We present a comprehensive empirical analysis across three computer vision tasks, 13 distinct attack types, and 10 training strategies, resulting in 130k individual evaluations.
    \item
    We present a comprehensive empirical analysis across three computer vision tasks, 13 distinct attack types, and 10 training strategies.
    Our ``\textbf{DUMB population}'' comprises 240 uniquely trained models spanning diverse datasets and architectures, resulting in over 130,000 individual evaluations.
    \item Our analysis provides novel insights and best practices on how to apply adversarial training to maximize model robustness under realistic attack scenarios involving transferability.
    \item We release the full codebase and evaluation scripts to promote reproducibility: 
    \url{https://github.com/spritz-group/DUMBer}.
\end{itemize}

% \paragraph{Findings.}
% After evaluating many evasion attacks on all possible combinations of adversarial training techniques, dataset source, model architecture, and class balance of the datasets, our findings can be summarized as follows.
% \begin{enumerate}
%     \item \fra{Insight \#1}
%     \item \fra{Insight \#2}
%     \item \fra{Insight \#3}\marco{adversarial training with non-math adv samples can be quite effective and independent from dumb cases}\fra{cazzata bro} \marco{ho cappato bro}
% \end{enumerate}

\paragraph{Research Questions.}
The extensive cross-parameter evaluations conducted in this study enable us to address unique research questions that are highly valuable to AI practitioners and future research efforts.
% \begin{enumerate}[label=\textbf{RQ\arabic*:}, leftmargin=*, align=left]
\begin{enumerate}\setlength{\itemindent}{1.8em}
    \item[\textbf{RQ1:}]
    Which adversarial attacks are most effective across the entire DUMB population?
    In other words, what attack strategies perform best regardless of the adversary's level of knowledge or access?
    \item[\textbf{RQ2:}]
    Which adversarial training strategies offer the highest overall robustness?
    What defense techniques are most effective across the diverse scenarios represented in the DUMB population, irrespective of the attacker's assumptions?
    \item[\textbf{RQ3:}]
    How do different training strategies perform across varying evaluation scenarios \texttt{C}?
    In other words, what are the best- and worst-case conditions for deploying each defense strategy in practice?
    \item[\textbf{RQ4:}]
    Do robust defenses against strong attacks generalize to weaker ones?
    Can adversarial training improve resilience uniformly across a spectrum of attack strengths, or is its benefit limited to specific threat levels?
    \item[\textbf{RQ5:}]
    When and where does adversarial training fail?
    Specifically, how frequently does it result in a negative AMR, and are these failures uniformly distributed across scenarios \texttt{C}?
\end{enumerate}

% \paragraph{Organization.}
% This paper is structured as follows.  
% Section~\ref{sec:related} reviews the state-of-the-art in adversarial machine learning and adversarial training.  
% In Section~\ref{sec:model}, we introduce the DUMBer attacker model, followed by a detailed description of our experimental methodology in Section~\ref{sec:methodology}.  
% Our empirical findings and key insights are presented in Section~\ref{sec:results}, while Section~\ref{sec:conclusions} concludes the paper and outlines directions for future work.
% \input{Sections/02_Background}
\section{Related Works}
\label{sec:related}

We now review the state-of-the-art in adversarial training and transfearbility.

\paragraph{Adversarial Transferability.}
% Adversarial examples are crafted to fool machine learning models and often transfer across different models, enabling black- or gray-box attacks using surrogate models.
% Recent studies have explored the mechanisms and factors influencing adversarial transferability.
% For instance, Gu et al.~\cite{gu2023survey} provide a comprehensive survey categorizing methodologies to enhance the transferability of adversarial examples, discussing fundamental principles and challenges in the field.
% Their work highlights the need for robust evaluation frameworks to assess the transferability across diverse model architectures and tasks.
% Building upon this, Yu et al.~\cite{yu2023reliable} emphasize the importance of reliable evaluation protocols, revealing that adversarial transferability is often overestimated when assessments are confined to similar model architectures, such as CNNs.
% They advocate for broader benchmarks covering diverse neural network architectures to more accurately assess adversarial example transferability.
% These insights underscore the necessity for frameworks like DUMB and DUMBer, which systematically evaluate adversarial transferability across multiple dimensions, including dataset sources, model architectures, and class balance.
Adversarial examples are designed to deceive machine learning models and often transfer across different models, enabling black- and gray-box attacks through surrogate models.
Recent work has investigated the factors behind adversarial transferability.
Gu et al.\cite{gu2023survey} provide a comprehensive survey, categorizing methods to enhance transferability and outlining core principles and challenges.
They stress the need for robust evaluation frameworks covering diverse architectures and tasks.
Building on this, Yu et al.\cite{yu2023reliable} show that transferability is often overestimated when evaluations are limited to similar architectures, such as CNNs, and call for broader benchmarks across different neural networks.
These findings motivate the need for frameworks like DUMB and DUMBer, which systematically assess adversarial transferability across dimensions like dataset source, model architecture, and class balance.

\paragraph{Adversarial Training.}
Adversarial training is a leading defense strategy that incorporates adversarial examples into the training process to improve model robustness.
Early approaches include FGSM-based training~\cite{goodfellow2014explaining}, which uses the Fast Gradient Sign Method, and PGD-based training~\cite{madry2017towards}, which applies iterative Projected Gradient Descent for stronger perturbations.
Both aim to defend against both seen and unseen attacks.
Curriculum adversarial training~\cite{cai2018curriculum} extends these ideas by gradually increasing perturbation strength ($\epsilon$) during training, promoting more stable robustness.
Ensemble adversarial training~\cite{tramer2017ensemble} further diversifies defenses by introducing adversarial examples from multiple pre-trained models, improving resilience against transfer attacks.
With DUMBer, we evaluate these techniques and their variants, also considering model-agnostic perturbations, to systematically measure their impact on robustness across diverse and transferable adversarial scenarios.
\section{Threat Model}
\label{sec:model}

The original DUMB attacker model emphasizes the importance of simulating realistic conditions for adversarial transferability—conditions often neglected in the current literature~\cite{grosse2023machine}.
Building on these foundations, our DUMBer framework explicitly addresses three critical challenges that arise during attack execution in practical settings:
\begin{itemize}
    \item \textit{Dataset:} Most prior works assume that both the attacker and victim have access to the same dataset, which is rarely the case in real-world scenarios.
    Constructing a surrogate dataset is far from trivial, as it depends heavily on corpus generation strategies that can vary significantly across domains and institutions.
    For example, in hate speech detection, it has been shown that existing datasets are constructed using divergent methodologies, leading to models that perform well on their training data but generalize poorly to others~\cite{grondahl2018all}.
    This undermines the assumption that transferability is guaranteed simply by using adversarial examples.
    \item \textit{Ground-truth distribution:} A related yet distinct issue concerns the assumption that attacker and victim datasets share the same underlying distribution.
    In practice, this is rarely true.
    Differences may stem from distinct data collection processes or disparate preprocessing and augmentation techniques.
    This mismatch becomes even more pronounced in imbalanced tasks, where techniques such as SMOTE~\cite{chawla2002smote} or GAN-based oversampling~\cite{frid2018synthetic} are often employed to rebalance class distributions, further distorting the comparability of training data across parties.
    \item \textit{Model architecture:} Attackers and victims typically do not rely on identical models.
    While prior work sometimes evaluates transferability across model families, the space of potential architectures is vast, ranging from standard CNNs to more sophisticated and customized models.
    For instance, in computer vision alone, one might choose among VGG variants (e.g., VGG16, VGG19) or ResNet families (e.g., ResNet18, ResNet50), each of which may respond differently to adversarial perturbations.
    This diversity introduces an additional layer of unpredictability in the effectiveness of transfer-based attacks.
\end{itemize}
In Table~\ref{tab:rainbow}, we summarize eight representative attack scenarios, each capturing a possible mismatch between the attacker’s surrogate model and the victim’s target model.
% The subscripts $a$ and $v$ denote attacker and victim, respectively.
In realistic settings, the attacker typically does not know which scenario they are operating in, except in the idealized white-box case.

\begin{table}[!htbp]
    \centering
    \caption{Conditions for each case in the DUMB attacker model. Scenarios for each \texttt{C} are included in~\cite{alecci2023dumb}. Subscripts $a$ and $v$ denote attacker and victim, respectively.}
    \label{tab:rainbow}
    \begin{tabular}{c|c|c|c}
        \toprule
        \textbf{Case} & \textbf{DU$_a$ vs DU$_v$} & \textbf{M$_a$ vs M$_v$} & \textbf{B$_a$ vs B$_v$} \\
        \midrule
        % \rowcolor{gray!15} \texttt{C1} & \ding{51} & \ding{51} & \ding{51} \\
        % \texttt{C2} & \ding{51} & \ding{51} & \ding{55} \\
        % \rowcolor{gray!15} \texttt{C3} & \ding{51} & \ding{55} & \ding{51} \\
        % \texttt{C4} & \ding{51} & \ding{55} & \ding{55} \\
        % \rowcolor{gray!15} \texttt{C5} & \ding{55} & \ding{51} & \ding{51} \\
        % \texttt{C6} & \ding{55} & \ding{51} & \ding{55} \\
        % \rowcolor{gray!15} \texttt{C7} & \ding{55} & \ding{55} & \ding{51} \\
        % \texttt{C8} & \ding{55} & \ding{55} & \ding{55} \\
        % \bottomrule
        % \multicolumn{4}{l}{\scriptsize{\ding{51} = match, \ding{55} = mismatch.}}
        \rowcolor{gray!15} \texttt{C1} & \fullcirc & \fullcirc & \fullcirc \\
        \texttt{C2} & \fullcirc & \fullcirc & \emptycirc \\
        \rowcolor{gray!15} \texttt{C3} & \fullcirc & \emptycirc & \fullcirc \\
        \texttt{C4} & \fullcirc & \emptycirc & \emptycirc \\
        \rowcolor{gray!15} \texttt{C5} & \emptycirc & \fullcirc & \fullcirc \\
        \texttt{C6} & \emptycirc & \fullcirc & \emptycirc \\
        \rowcolor{gray!15} \texttt{C7} & \emptycirc & \emptycirc & \fullcirc \\
        \texttt{C8} & \emptycirc & \emptycirc & \emptycirc \\
        \bottomrule
        \multicolumn{4}{l}{\scriptsize{\fullcirc\:= match, \emptycirc\:= mismatch.}}\\
        \multicolumn{4}{l}{\scriptsize{\texttt{C1} = pure white-box, \texttt{C8} = pure black-box}}\\
    \end{tabular}
\end{table}

Beyond these structural mismatches, an additional layer of complexity arises from the widespread adoption of adversarial training techniques.
Deployed models, especially those exposed to end-users, are often hardened through such defenses to enhance robustness and safety alignment.
As a result, attackers must overcome transferability challenges and contend with models explicitly trained to resist adversarial inputs.
The objective of DUMBer is to systematically evaluate how these real-world conditions, ranging from mismatched assumptions to adversarial training, impact the effectiveness of transfer-based attacks.
\section{Methodology}
\label{sec:methodology}

\begin{figure}
    \centering
    \includegraphics[width=.75\linewidth]{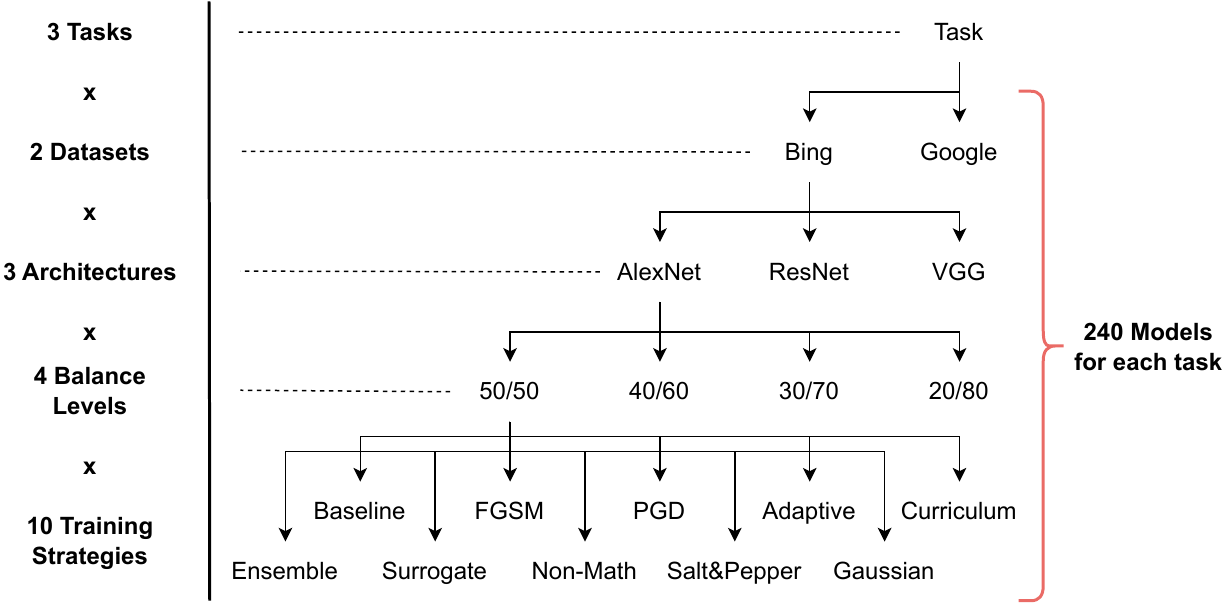}
    \caption{Model combinations during the training phase.}
    \label{fig:combinations}
\end{figure}

% We now detail our methodology.
% While our evaluation of transferability focuses on the eight specific cases introduced in Section~\ref{sec:model}, our primary contribution lies in the systematic analysis of how adversarial training methodologies impact those scenarios.
% To this end, in Section~\ref{subsec:transferability_dimensions}, we define the dimensions relevant to adversarial transferability; in Section~\ref{subsec:attacks}, we overview the adversarial attacks under consideration; in Section~\ref{subsec:training_dimensions}, we introduce the dimensions of adversarial training we consider; and finally, in Section~\ref{subsec:testing}, we present our testing methodology, which brings these components together into a unified evaluation framework.\footnote{Each underlined element in this section represents a key dimension that influences either the DUMB population or the evaluation setup.}
% An overview of our training methodology is shown in Fig.~\ref{fig:combinations}.
Next, we present our methodology, which is built on the foundation of the DUMB framework.
While our evaluation covers the eight cases from Section~\ref{sec:model}, our main focus is a systematic study of how adversarial training affects these scenarios.
Section~\ref{subsec:transferability_dimensions} defines the dimensions of transferability, Section~\ref{subsec:attacks} overviews the attacks, Section~\ref{subsec:training_dimensions} describes training dimensions, and Section~\ref{subsec:testing} explains our unified testing framework.\footnote{Underlined terms highlight key dimensions impacting the DUMB population or evaluation setup.}
An overview of the training process is shown in Fig.~\ref{fig:combinations}.

\subsection{Transferability Dimensions}
\label{subsec:transferability_dimensions}

This work focuses on three distinct computer vision tasks, each representing common settings found in adversarial attack literature.
These tasks are framed as binary classification problems: distinguishing between Bikes vs. Motorbikes (\underline{B\&M}), Cats vs. Dogs (\underline{C\&D}), and Men vs. Women (\underline{M\&W}).
For each task, we systematically vary the dimensions that influence adversarial transferability.

\subsubsection{Dataset Source}
To meet the specific needs of our testbed, we manually collect and validate two distinct datasets for each task from \underline{Bing} and \underline{Google}, ensuring control over complexity and potential biases.
Starting with an average of 14,264 images per dataset, we remove duplicates using difPy, perform manual inspections to eliminate mislabeled or low-quality samples, and randomly select 10,000 balanced images per dataset.
All images are then resized to 300×300 RGB using anti-aliasing to maintain quality.

\subsubsection{Model Architecture}
We employ three widely-used and well-established computer vision architectures: \underline{AlexNet}~\cite{krizhevsky2014one}, \underline{ResNet18}~\cite{he2016deep}, and \underline{VGG11}~\cite{simonyan2014very}.
We fine-tune these models across our experimental tasks, building on architectures also explored in the DUMB framework.
Training procedures align with the official PyTorch guidelines to ensure reproducibility.

\subsubsection{Ground Truth Balancing}
To assess the impact of class imbalance between attacker and defender, we simulate four levels of imbalance in the training sets: balanced (\underline{50/50}), weak (\underline{40/60}), medium (\underline{30/70}), and strong (\underline{20/80}), always treating the first class (Cats, Men, Bikes) as the minority across all tasks.
The minority class is undersampled accordingly while keeping the majority class fixed (e.g., for strong imbalance in the Cats vs. Dogs task, 875 Cats vs. 3500 Dogs).
This procedure affects only the training sets; validation and test sets remain balanced.

\subsection{Attacks}
\label{subsec:attacks}

We incorporate two broad categories of adversarial attacks, mathematical and non-mathematical, for a total of 13 distinct attacks across different threat models.

\subsubsection{Mathematical Attacks}
Mathematical attacks are model-aware and generated through optimization methods that leverage gradient information.
Our evaluation includes widely used attacks such as \underline{FGSM}~\cite{goodfellow2014explaining}, \underline{BIM}~\cite{kurakin2018adversarial}, \underline{PGD}~\cite{madry2017towards}, \underline{RFGSM}~\cite{tramer2017ensemble}, \underline{DeepFool}~\cite{moosavi2016deepfool}, \underline{TIFGSM}~\cite{dong2019evading}, and \underline{Square}~\cite{andriushchenko2020square}, all implemented via the Torchattacks Python library.
These attacks differ in complexity and transferability, offering a robust basis to test model vulnerability under white-box and black-box scenarios.
It is important to note that mathematical attacks are generated at the “balance level” described in Fig.~\ref{fig:combinations}, without incorporating any adversarial training on the source model used for attack generation.

\subsubsection{Non-Mathematical Attacks}
On the other hand, non-mathematical attacks rely on simple image transformations independent of any model, making them practical in real-world settings.
Using only basic image processing via PIL, we simulate attacks like \underline{Gaussian noise}, \underline{grayscale} conversion, \underline{box blur}, \underline{salt-and-pepper} noise, \underline{random occlusion} (black box), and \underline{color inversion}.
Though simpler, these attacks pose realistic threats to vision models and highlight the relevance of evaluating robustness beyond optimization-based techniques.

\subsubsection{Parameter Tuning}
% Each attack, whether mathematical or non-mathematical, includes at least one parameter that controls the strength or intensity of the perturbation.
% For example, most mathematical attacks rely on an epsilon ($\epsilon$) value.
% Similarly, Gaussian noise, blur, or random occlusion transformations include parameters like radius, noise level, or occlusion size.
% To ensure practical and visually realistic perturbations, we tune parameters by maximizing the attack success while maintaining perceptual similarity between original and adversarial samples, measured via the Structural Similarity Index Measure (SSIM).
% A minimum SSIM threshold of 0.4 is enforced to avoid overly distorted inputs.
% This tuning process is performed on the test set, as these attacks are used to evaluate model robustness in a pure evasion setting.
% We also apply the same attack generation process to the validation set with a different objective.
% Instead of selecting a single optimal parameter, we retain all perturbed samples across a range of parameter values (e.g., multiple $\epsilon$ levels).
% These pre-generated adversarial examples from the validation set support specific adversarial training strategies explored in our study.
% This separation ensures we do not leak information between the (adversarial) training and test sets, maintaining the integrity of our evaluation.
Each attack, whether mathematical or not, has at least one parameter controlling its strength (e.g., $\epsilon$ for mathematical attacks, radius or noise level for others).
We tune these parameters to maximize attack success while preserving visual similarity, enforcing a minimum Structural Similarity Index Measure (SSIM) of 0.4 to prevent excessive distortion.
For testing, we select optimal parameters to evaluate model robustness under pure evasion.
For validation, we generate adversarial examples across multiple parameter values to support adversarial training, ensuring no information leakage between training and testing.

\subsection{Training Strategies}
\label{subsec:training_dimensions}

We begin by training a baseline model on unperturbed data, reflecting nominal conditions without adversarial influence.
This is a control scenario to assess the degradation caused by adversarial attacks.
Subsequently, we investigate nine adversarial training strategies designed to enhance robustness against evasion attacks.
All adversarial training experiments are conducted from scratch rather than fine-tuning a nominally trained model.
This choice avoids compromising between adversarial and nominal performance—an issue commonly encountered in fine-tuning approaches.
For each adversarial training strategy, we construct a training dataset composed of 80\% original (clean) samples and 20\% adversarial samples.
These adversarial examples are drawn from the validation set, following the generation procedure described in Section~\ref{subsec:attacks}.
While this general setup holds for all strategies, specific implementations may vary based on the training logic.
We define and adapt the following training strategies to facilitate integration within our framework.
\begin{itemize}
    \item \underline{FGSM:} Includes validation-time adversarial samples generated using FGSM at a fixed $\epsilon = 0.2$, selected to balance perturbation visibility and attack strength~\cite{goodfellow2014explaining}.
    \item \underline{PGD:} Mirrors the FGSM setup but employs PGD as the attack method, also with $\epsilon = 0.2$~\cite{madry2017towards}.
    \item \underline{Ensemble:} Incorporates adversarial examples from both FGSM and PGD attacks, aiming to increase robustness through attack diversity~\cite{tramer2017ensemble}.
    \item \underline{Surrogate:} Similar to Ensemble, but uses adversarial examples generated by different model architectures (trained under the same DUMB configuration) to simulate transferability-based threats.
    \item \underline{Curriculum:} Progressively increases $\epsilon$ over training epochs while generating FGSM attacks offline~\cite{cai2018curriculum}.
    This method introduces gradually harder adversarial examples during training.
    \item \underline{Adaptive:} Also increases $\epsilon$ over time but generates adversarial examples online during training~\cite{marchiori2024canederli}.
    This approach adapts to the model’s current weaknesses, simulating evolving attack sophistication.
    \item \underline{Non-Mathematical Mixture:} Employs all non-mathematical attacks at fixed perturbation strengths, ensuring diversity without reliance on model gradients.
    \item \underline{Gaussian:} Uses only Gaussian noise as the adversarial perturbation in the 20\% adversarial portion of the training set.
    \item \underline{Salt-and-Pepper:} Uses only salt-and-pepper noise to generate adversarial examples for training, enabling focused analysis of noise-based robustness.
\end{itemize}

\subsection{Testing}
\label{subsec:testing}

Designing the evaluation framework presents unique challenges, as the complexity of adversarial attack generation is not aligned with the training process of the selected target model, unlike the original DUMB framework, where these components are tightly coupled.

\subsubsection{Mathematical Attacks}
Gradient-based adversarial attacks must be crafted using dedicated models, referred to as \textit{source models} $M_{\text{src}}$.
Our transferability dimensions determine the number of these models: for each task, we consider 2 dataset sources, 3 model architectures, and 4 data balance levels, resulting in $2 \times 3 \times 4 = 24$ source models per task.
Each of the 7 mathematical attacks is then evaluated against a broader set of \textit{target models} $M_{\text{trg}}$, which includes the full combination space shown in Fig.~\ref{fig:combinations}, amounting to 240 unique configurations.
Consequently, the total number of mathematical attack evaluations amounts to: 3 tasks $\times$ 7 attacks $\times$ 24 $M_{src}$ $\times$ 240 $M_{trg}$ = 120,960, where 240 $M_{trg}$ correspond to 2 dataset sources $\times$ 3 model architectures $\times$ 4 data balance levels $\times$ 10 training strategies.

\subsubsection{Non-Mathematical Attacks}
Model-agnostic image perturbations, such as those not relying on gradients, do not require dedicated $M_{\text{src}}$ and can be applied directly to the input data.
Nonetheless, the dataset source remains relevant, as the images differ between the Bing and Google datasets.
As a result, the total number of evaluations for non-mathematical attacks is: 3 tasks $\times$ 6 attacks $\times$ 2 datasets $\times$ 240 $M_{trg}$ = 8,640.
\section{Results}
\label{sec:results}

In this section, we present the results of our evaluation framework and the effect that adversarial training strategies have on the different transferability cases.
We first define the metrics for our evaluation in Section~\ref{subsec:metrics}, followed by a baseline evaluation of our trained models in Section~\ref{subsec:baseline}.
% We then aim to answer the following research questions.
We then answer the research questions detailed in Section~\ref{sec:introduction}.

\subsection{Metrics}
\label{subsec:metrics}

From a machine learning standpoint, each of our tasks is formulated as a binary classification problem, where the model learns to distinguish between two distinct classes based on the input images.
Since label balancing plays a central role in our analysis, we adopt the F1 score as the primary evaluation metric to assess the classification performance of the models.
The F1 score provides a balanced measure of precision and recall, making it particularly suitable for evaluating performance in scenarios where class distribution may vary.
This score is defined as follows.
\begin{equation}
    F1 = 2\frac{precision \cdot recall}{precision + recall}.
\end{equation}

To assess both the impact of adversarial attacks and the robustness of adversarially trained models, we introduce two additional evaluation metrics: Attack Success Rate (ASR) and Attack Mitigation Rate (AMR).
ASR, also used in the original DUMB framework, measures the proportion of samples classified initially correctly by a model under clean conditions but misclassified after applying the attack.
This quantifies the effectiveness of the attack in degrading model performance.
To enable a more fine-grained assessment, we also introduce a metric called \textit{severity}, assigning each attack attempt to one of five levels that evenly partition the ASR range from 0\% to 100\%.
This distinction captures attacks ranging from minimal impact (severity score 1) to significant degradation (severity score 5).
AMR, instead, is the difference between the ASR before and after adversarial training, normalized by the ASR before training.
Specifically, AMR is calculated as:
\begin{equation}
    AMR = \frac{ASR_{original} - ASR_{adv}}{ASR_{original}},
\end{equation}
% \fra{mettiamo il logaritmo altrimenti anche da 6\% a 3\% vuol dire 50\% di improvement che meh stiamo flexando troppo}
where the subscript ``original'' refers to the ASR measured on the baseline model, and ``adv'' refers to the ASR of the adversarially trained model.
This metric quantifies the attack's success rate reduction following adversarial training.
A higher AMR indicates that a specific training strategy has more effectively mitigated the attack’s impact, improving the model’s robustness.
It is important to note that AMR is upper-bounded at 100\% (indicating that all attacks were mitigated successfully), but it is theoretically not lower-bounded.
This is because adversarial training could also deteriorate the model’s performance in nominal and adversarial scenarios, resulting in a negative or zero value for AMR.
To maintain interpretability, we cap AMR values at -100\%, acknowledging that stronger degradations are possible, though they often occur when the original ASR was already low.
We thus define +100\% as perfect improvement and -100\% as complete degradation.
% \marco{theoretically -inf se ASR di partenza é zero. Possiamo dire che cappiamo a -100. o NO CAP???}

\subsection{Baseline Evaluation}
\label{subsec:baseline}

Before assessing the impact of evasion attacks, it is crucial to ensure that our models perform reliably under standard, unperturbed conditions.
Table~\ref{tab:baseline} presents the F1 scores across all models and training strategies on the nominal dataset.
As we adopt the same tasks defined in the original DUMB framework, we observe a consistent trend in the baseline evaluation, where B\&M emerges as the most straightforward task, while M\&W proves to be the most challenging.
Overall, adversarial training does not significantly degrade performance on clean data.
Most strategies maintain parity with the baseline and, in some cases, even improve it.
Notably, Gaussian noise and Adaptive training slightly enhance model performance, suggesting a potential regularization effect that helps generalization.
The only approach that noticeably reduces performance is the Ensemble method, which shows an average drop of 4.26\% compared to the baseline.
This may be due to the added complexity of combining multiple decision boundaries, which could reduce precision in non-adversarial contexts.

\begin{table}[!htbp]
    \centering
    \caption{Model performance (F1 score) across different training strategies. Results are averaged on the dataset source and ground-truth balance dimensions.}
    \label{tab:baseline}
    \resizebox{\textwidth}{!}{
    \begin{tabular}{l|l|c|ccccccccc}
    \toprule
    \textbf{Task} & \textbf{Model} & \textbf{Base} & \textbf{FGSM} & \textbf{PGD} & \textbf{Ens.} & \textbf{Sur.} & \textbf{Cur.} & \textbf{Ada.} & \textbf{N.M.} & \textbf{Gau.} & \textbf{S.\&P.} \\
    \midrule
    \multirow{3}{*}{B\&M} & AlexNet & 0.976 & 0.978 & 0.979 & 0.892 & 0.979 & 0.978 & 0.978 & 0.979 & 0.980 & 0.977 \\ 
    & ResNet & 0.986 & 0.987 & 0.986 & 0.987 & 0.988 & 0.987 & 0.985 & 0.987 & 0.987 & 0.988 \\ 
    & VGG & 0.985 & 0.985 & 0.986 & 0.987 & 0.988 & 0.986 & 0.985 & 0.987 & 0.985 & 0.986 \\
    \midrule
    \multirow{3}{*}{C\&D} & AlexNet & 0.953 & 0.948 & 0.949 & 0.933 & 0.947 & 0.952 & 0.953 & 0.951 & 0.950 & 0.948 \\ 
    & ResNet & 0.978 & 0.978 & 0.979 & 0.938 & 0.978 & 0.978 & 0.978 & 0.978 & 0.979 & 0.978 \\ 
    & VGG & 0.982 & 0.981 & 0.982 & 0.939 & 0.981 & 0.981 & 0.982 & 0.980 & 0.982 & 0.981 \\
    \midrule
    \multirow{3}{*}{M\&W} & AlexNet & 0.858 & 0.862 & 0.861 & 0.825 & 0.857 & 0.856 & 0.868 & 0.864 & 0.865 & 0.859 \\ 
    & ResNet & 0.921 & 0.921 & 0.922 & 0.871 & 0.917 & 0.925 & 0.920 & 0.920 & 0.922 & 0.922 \\ 
    & VGG & 0.925 & 0.925 & 0.925 & 0.831 & 0.923 & 0.925 & 0.925 & 0.920 & 0.925 & 0.921 \\
    \midrule
    \midrule
    \multicolumn{3}{l|}{Avg. Change w.r.t Base} 
        & \cellcolor{Green!15}+0.02\% 
        & \cellcolor{Green!15}+0.06\% 
        & \cellcolor{Red!15}-4.26\% 
        & \cellcolor{Red!15}-0.08\% 
        & \cellcolor{Green!15}+0.05\% 
        & \cellcolor{Green!15}+0.13\% 
        & \cellcolor{Green!15}+0.03\% 
        & \cellcolor{Green!15}+0.14\% 
        & \cellcolor{Red!15}-0.05\% \\
    \bottomrule
    \multicolumn{12}{l}{\scriptsize{\textbf{Ens.} = Ensemble, \textbf{Sur.} = Surrogate, \textbf{Cur.} = Curriculum, \textbf{Ada.} = Adaptive.}} \\
    \multicolumn{12}{l}{\scriptsize{\textbf{N.M.} = Non-Math-Mix, \textbf{Gau.} = Gaussian noise, \textbf{S.\&P.} = Salt-and-pepper noise.}}
    \end{tabular}
    }
\end{table}

\subsection{Attack Overview}
\label{subsec:rq1}

\begin{figure}
    \centering
    \includegraphics[width=\linewidth]{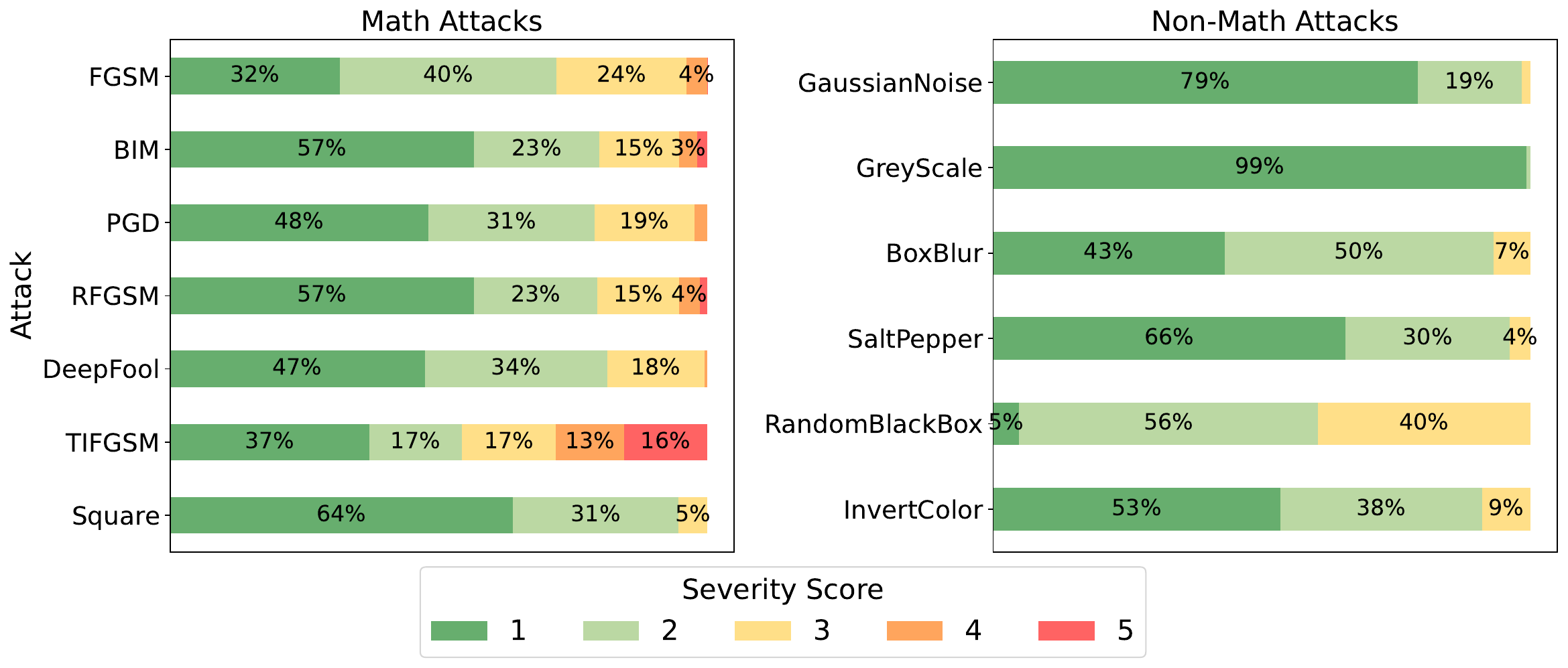}
    \caption{Severity score distribution of each attack.}
    \label{fig:asr_severity}
\end{figure}

Before analyzing the performance of the DUMB population across its subgroups, we first evaluate the overall effectiveness of the attacks described in Section~\ref{subsec:attacks}.
% To enable a more fine-grained assessment, we introduce a metric called \textit{severity}, assigning each attack attempt to one of five levels that evenly partition the ASR range from 0\% to 100\%.
% This distinction captures attacks ranging from minimal impact (severity score 1) to significant degradation (severity score 5).
An overview is presented in Fig.~\ref{fig:asr_severity}, where results are averaged across \texttt{C} scenarios and tasks to reflect general attack behavior under diverse, balanced conditions.
A more detailed scenario-specific analysis was already provided in the original DUMB paper~\cite{alecci2023dumb}.
Fig.~\ref{fig:asr_severity} shows that TIFGSM is the most effective strategy in the average DUMB scenario.
This is unsurprising given its design to enhance transferability, particularly benefiting ``grayer'' box settings.
Conversely, attacks such as BIM, Square, and RFGSM predominantly fall into severity score 1, suggesting limited effectiveness in more generalized or transfer-based contexts.
Non-mathematical attacks predictably show lower severity, although techniques like RandomBlackBox and BoxBlur occasionally outperform mathematical attacks such as BIM and Square, especially when averaging across all attacker scenarios.
Notably, within severity score 1, certain attacks display a large fraction of extremely low ASR cases: for example, BIM and RFGSM fall below a 5\% ASR in 24\% of evaluations, highlighting their limited transferability.
A more detailed analysis on each task is shown in Appendix~\ref{app:rq1}.
% \fra{dire che see ti interessa vedere per C, leggiti dumb1}
% \fra{citare che un tot è sotto il 5\% di asr, probabilmente figo da metttere in appendix}
% \fra{qualche parola su nonmath}

% \begin{formal}
% \textbf{RQ1:}
% \textit{Transfer-based attacks like TIFGSM are the most effective across the DUMB population, while non-mathematical attacks occasionally outperform traditional methods in specific scenarios.}
% \end{formal}

% \footnotetext{Values under 3\% are not labeled, but still displayed.}

\subsection{Adversarial Training Overview}
\label{subsec:rq2}

\begin{figure}
    \centering
    \includegraphics[width=\linewidth]{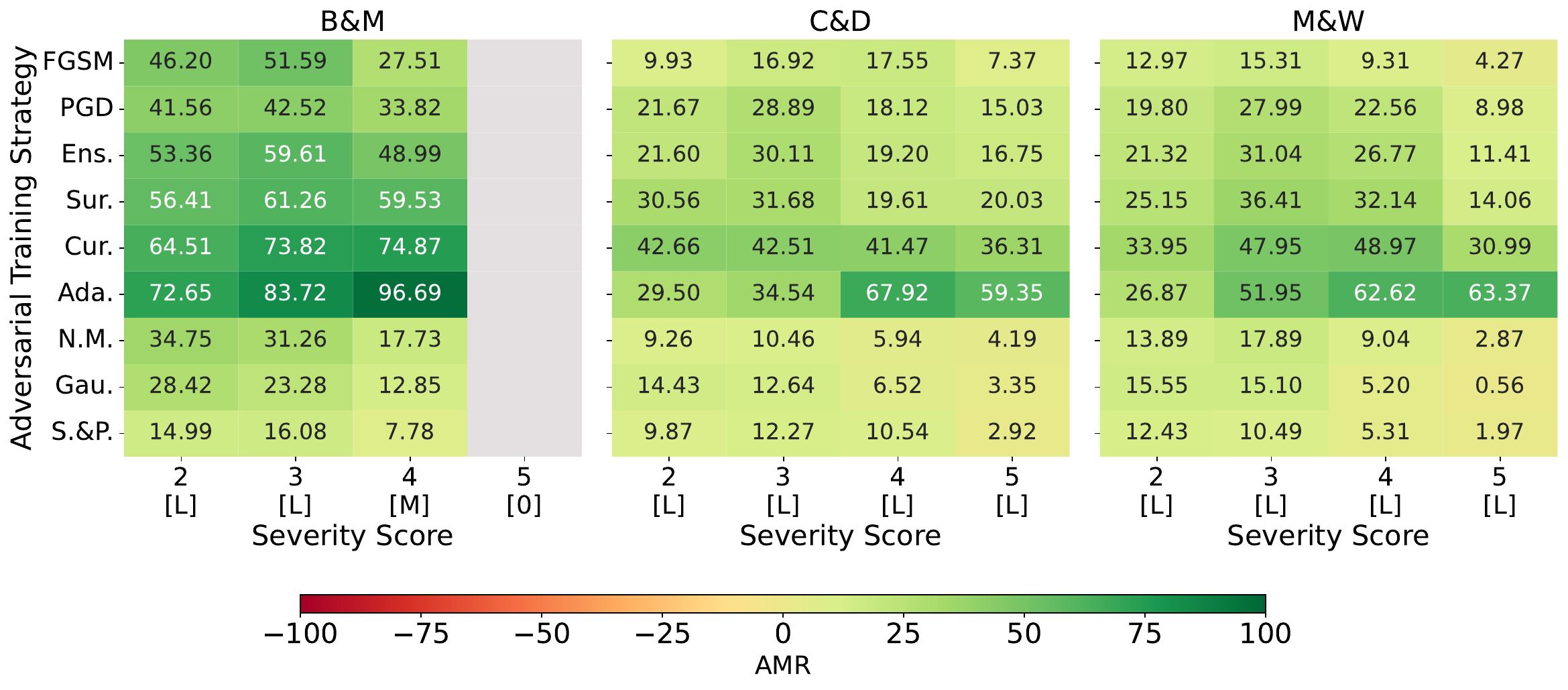}
    \caption{Training strategies' AMR at different attack severity scores. Labels below scores indicate sample sizes for each cell: \texttt{[S]} (1–10), \texttt{[M]} (10–50), and \texttt{[L]} (50+).}
    \label{fig:amr_heatmap}
\end{figure}

Following the approach in Section~\ref{subsec:rq1}, we now assess the AMR of each training strategy across the full DUMB population.
The goal is to identify which method performs best when the defender has no knowledge of the attacker's capabilities.
This evaluation considers each \texttt{C} scenario to reflect how a ``naive'' defender might deploy protection without tailored assumptions. Results are shown in Fig.~\ref{fig:amr_heatmap}.
Severity 5 is missing for B\&M, as no attacks reached that level.
Severity 1 is excluded across all tasks, as such attacks have negligible ASR and would distort AMR interpretation by exaggerating insignificant changes.
Our focus on higher severities aligns with the paper’s aim to support resilience against impactful threats. \textit{Adaptive} training stands out, achieving up to 96.69\% AMR on B\&M.
\textit{Curriculum} and \textit{surrogate} also perform well, while non-mathematical approaches offer modest but more consistent improvements.
Some negative AMR values appear, especially on the M\&W task. 
This stems from the task’s lower baseline F1 score (see Table~\ref{tab:baseline}), making it more sensitive to training disruptions.
Still, since AMR remains strongly positive at higher severities, practitioners should weigh these occasional drops—often linked to low-impact attacks—against the broader gains in robustness.
% \fra{bisogna dire che quando livello è 1 l'asr è pikkolo e quindi omettiamo}

% \begin{formal}
% \textbf{RQ2:}
% \textit{Adaptive adversarial training offers the highest overall robustness, with non-mathematical methods providing more consistent but modest resilience across diverse attack scenarios.}
% \end{formal}

\subsection{Adversarial Training on DUMB}
\label{subsec:rq3}

\begin{figure}
    \centering
    \includegraphics[width=\linewidth]{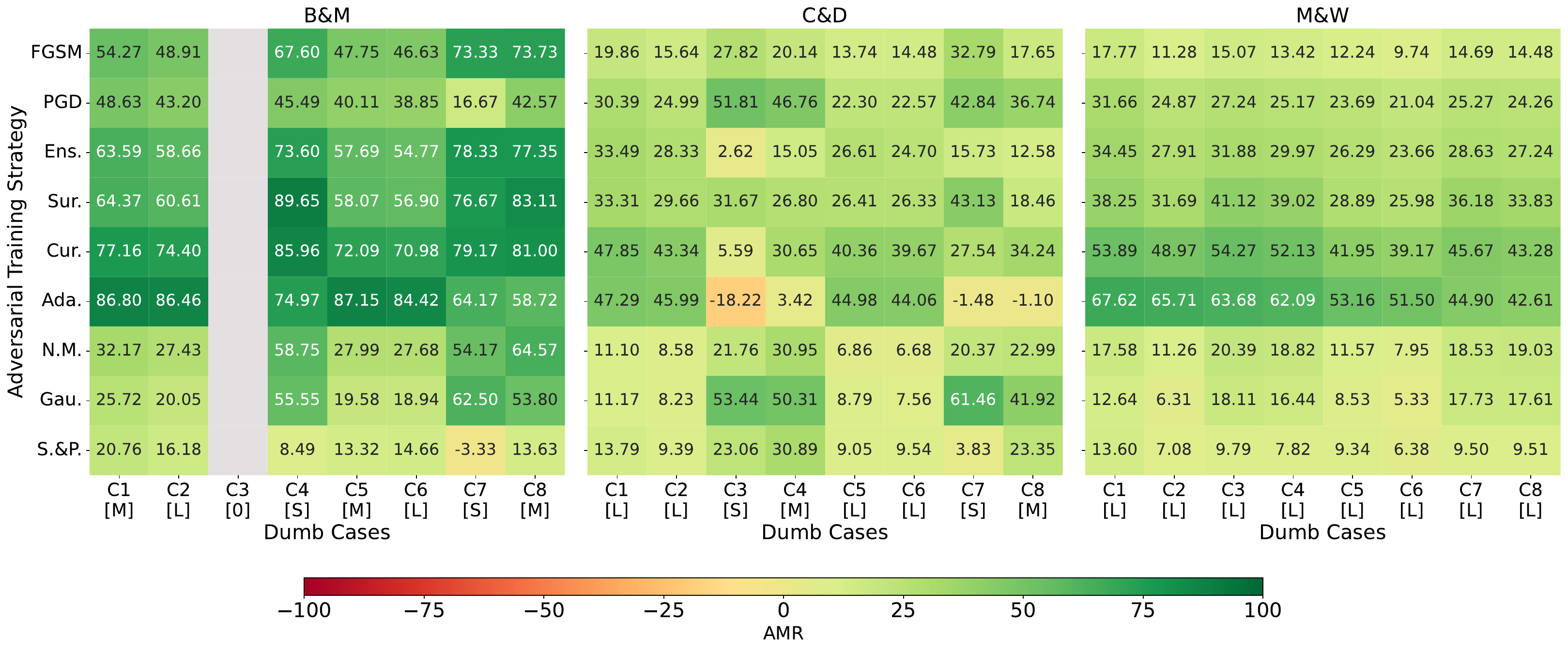}
    \caption{Training strategies' AMR at different DUMB cases. Labels below scores indicate sample sizes for each cell: \texttt{[S]} (1–10), \texttt{[M]} (10–50), and \texttt{[L]} (50+).}
    \label{fig:amrDumbCasesPerDefence}
\end{figure}

In Sections~\ref{subsec:rq1} and~\ref{subsec:rq2}, we provided general overviews of attacks and training strategies by averaging across all DUMB scenarios.
We now focus on a scenario-specific analysis, examining how each strategy performs under different \texttt{C} settings. 
This detailed view is intended for practitioners familiar with their threat model and the critical attack scenarios they face, as discussed in the original DUMB paper.
To ensure relevance, we limit our analysis to attacks with severity scores of 3 or higher.
This filters out cases where negative AMR values reflect negligible performance changes and emphasizes scenarios where adversarial training meaningfully impacts robustness.
These results are summarized in Fig.~\ref{fig:amrDumbCasesPerDefence}.
We observe that adversarial training is most effective when the source and target models share the same architecture (e.g., \texttt{C1}, \texttt{C2}, \texttt{C5}, \texttt{C8}), likely due to higher ASR in those cases, as noted in the original DUMB study.
Consistent with Section~\ref{subsec:rq2}, \textit{adaptive} training can yield negative AMR when model architectures differ, possibly due to lower initial ASR.
A noteworthy trend in the M\&W task is that AMR drops significantly when the dataset source differs, a pattern less evident in other tasks. This aligns with prior findings identifying M\&W as particularly sensitive to dataset mismatch, resulting in larger ASR discrepancies.
Finally, some task-scenario combinations are under-represented due to our severity $\geq$3 filter, especially where data is already sparse (e.g., \texttt{C1-4-5-7} in B\&M, \texttt{C3-4-7-8} in C\&D).
As such, outliers like the \textit{adaptive} \texttt{C3} case in C\&D should not be over-interpreted; broader evaluation is needed to confirm such trends.
% \fra{discussione di support con il numero dei samples}

% \fra{so che stiamo cambiando idea mille volte ma qua forse conviene mettere ogni task, stavo per scrivere le cose che avevamo visto in call (ossia dell'importanza di C3-4), ma da qua non si vede nulla.}

% \begin{figure}
%     \centering
%     \includegraphics[width=0.5\linewidth]{Images/Plots/RQ3_heatmapAdvxDumb_AMR_CrossTask.pdf}
%     \caption{AMR crosstalk per dumbCases \fra{Per Marco: to fix ordine delle adv tr plz} \marco{zio pera bro...}}
%     \label{fig:amrDumbCasesPerDefence}
% \end{figure}

\subsection{Attacks vs. Adversarial Training}
\label{subsec:rq4}

\begin{figure}[htbp]
    \centering
    \begin{subfigure}{0.475\linewidth}
        \centering
        \includegraphics[width=\linewidth]{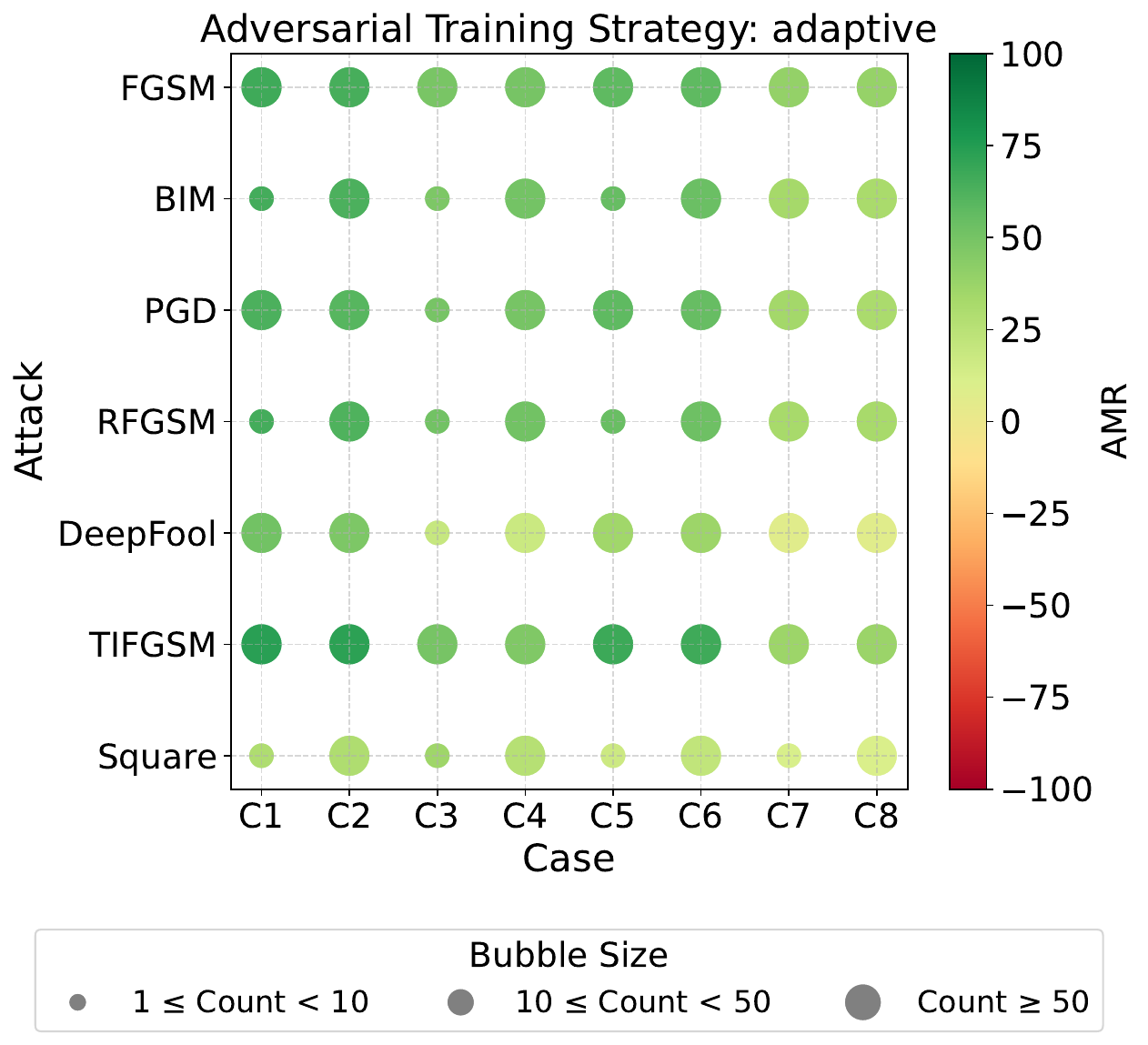}
        \caption{DUMB cases analysis.}
        \label{subfig:amr_ada_heatmaps_2x2_dumb}
    \end{subfigure}
    \hfill
    \begin{subfigure}{0.475\linewidth}
        \centering
        \includegraphics[width=\linewidth]{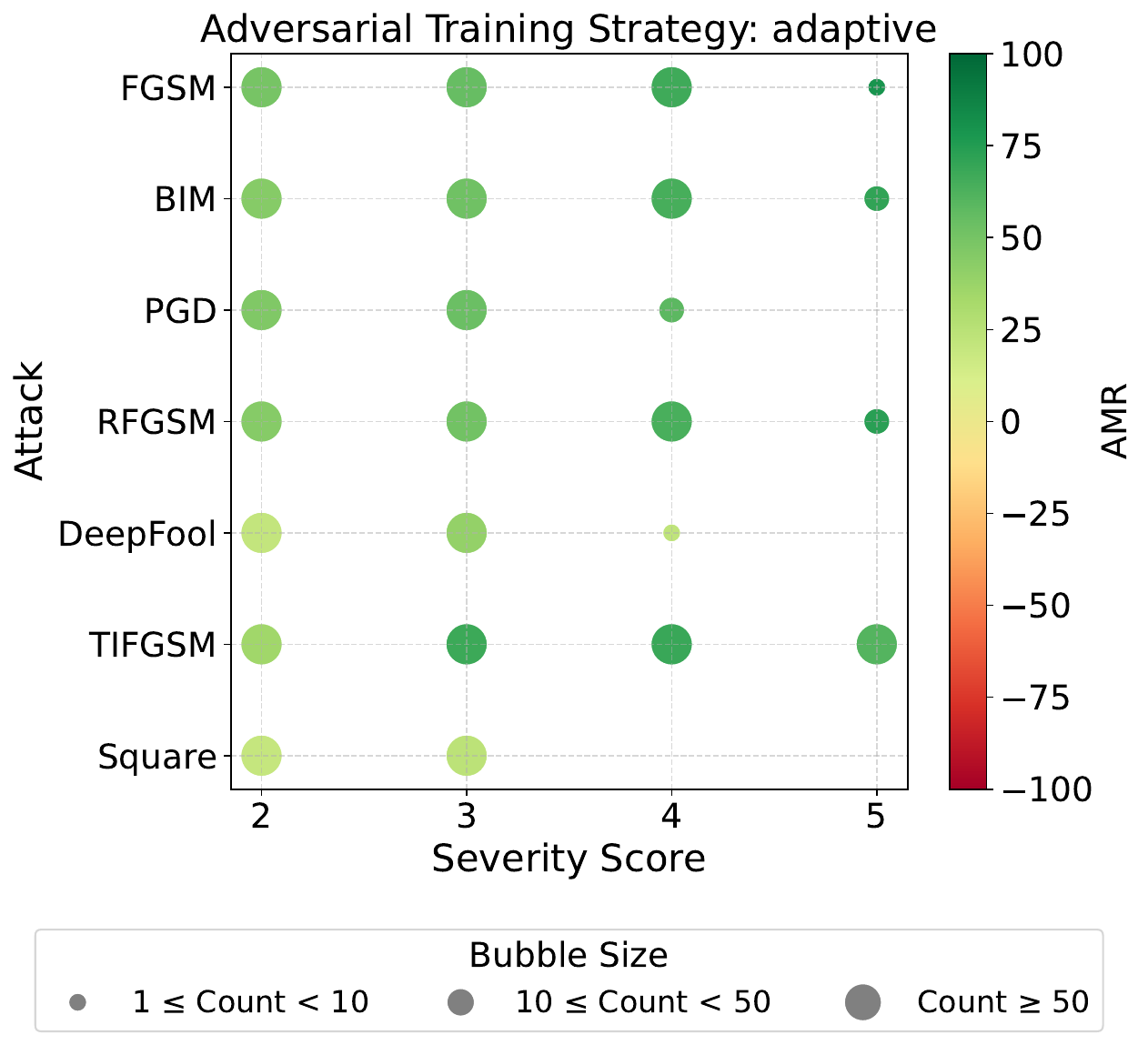}
        \caption{Attack severity score analysis.}
        \label{subfig:amr_ada_heatmaps_2x2_sev}
    \end{subfigure}
    % \vskip\baselineskip
    \caption{\textit{Adaptive} training strategy AMR under different attacks.}
    \label{fig:amr_ada_heatmaps_2x2}
\end{figure}

An ideal defense would maintain strong protection regardless of an attack’s severity.
However, adversarial training often involves trade-offs, where optimizing for one threat type can weaken resilience to others.
To explore this, we analyze two representative training strategies across varying attacks: \textit{adaptive}, a strong defense particularly effective against high-severity attacks (Fig.~\ref{fig:amr_ada_heatmaps_2x2}), and \textit{Non-Math}, a simpler, model-agnostic approach (Fig.~\ref{fig:amr_nonmath_heatmaps_2x2}).
A detailed overview of the specific AMR values is provided in Appendix~\ref{app:rq4}.
Examining their behavior across DUMB scenarios and attack severities reveals clear patterns.
\textit{Adaptive} training offers strong resilience against high-severity attacks, often achieving AMR values above 60\% for severities 4 and 5.
However, its performance drops against lower severities, where most negative AMR values occur.
Conversely, \textit{Non-Math} achieves more consistent but modest positive AMR across lower severities, though it struggles against stronger attacks.
Neither method achieves uniform robustness: \textit{adaptive} favors severe threats, while \textit{Non-Math} provides broader but shallower protection.
This highlights a key trade-off for practitioners: optimizing for strong attacks may expose vulnerabilities to weaker but more frequent perturbations.

% \vspace{-15mm}
\begin{figure}[htbp]
    \centering
    \begin{subfigure}{0.475\linewidth}
        \centering
        \includegraphics[width=\linewidth]{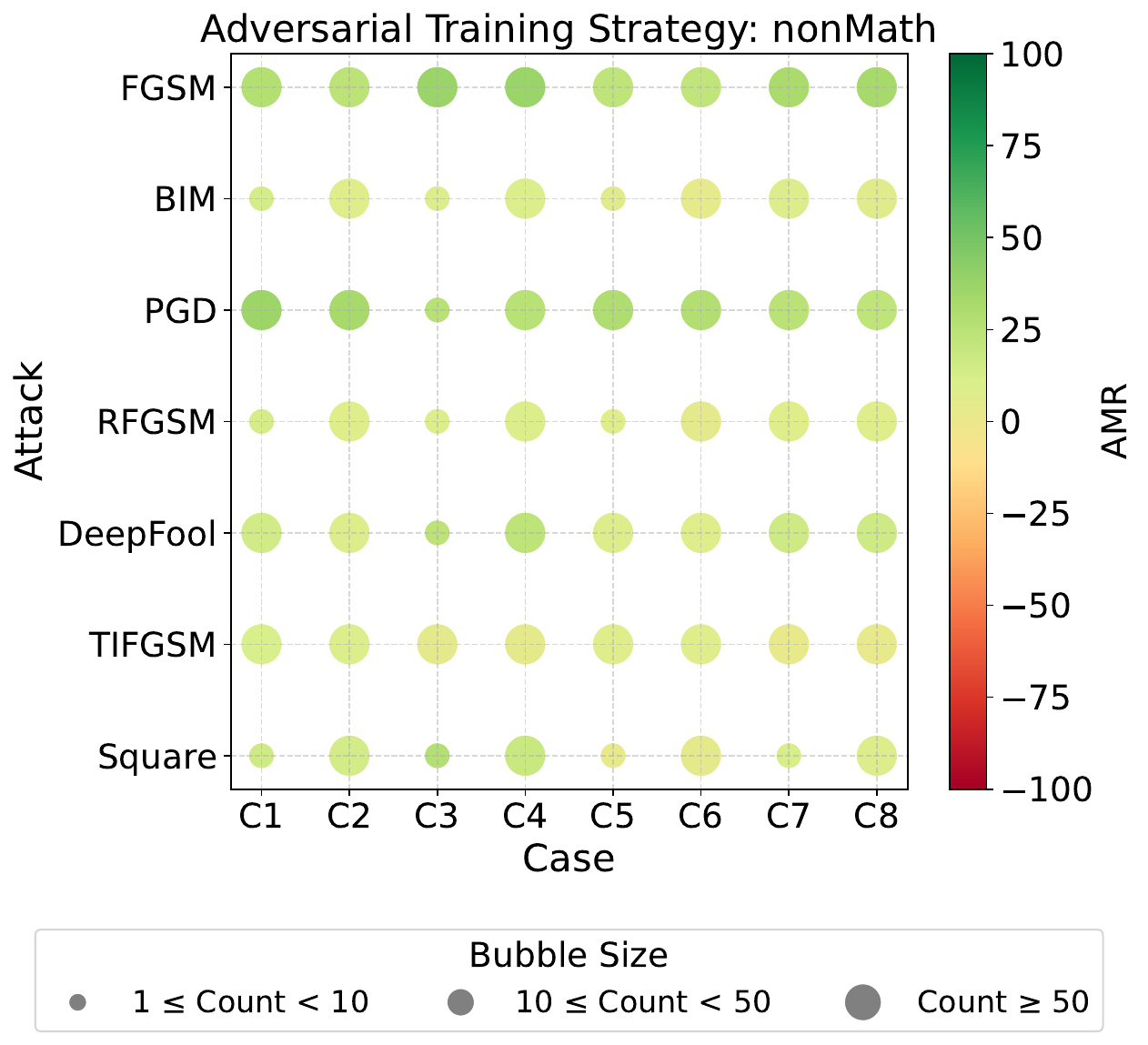}
        \caption{DUMB cases analysis.}
        \label{subfig:amr_nonmath_heatmaps_2x2_dumb}
    \end{subfigure}
    \hfill
    \begin{subfigure}{0.475\linewidth}
        \centering
        \includegraphics[width=\linewidth]{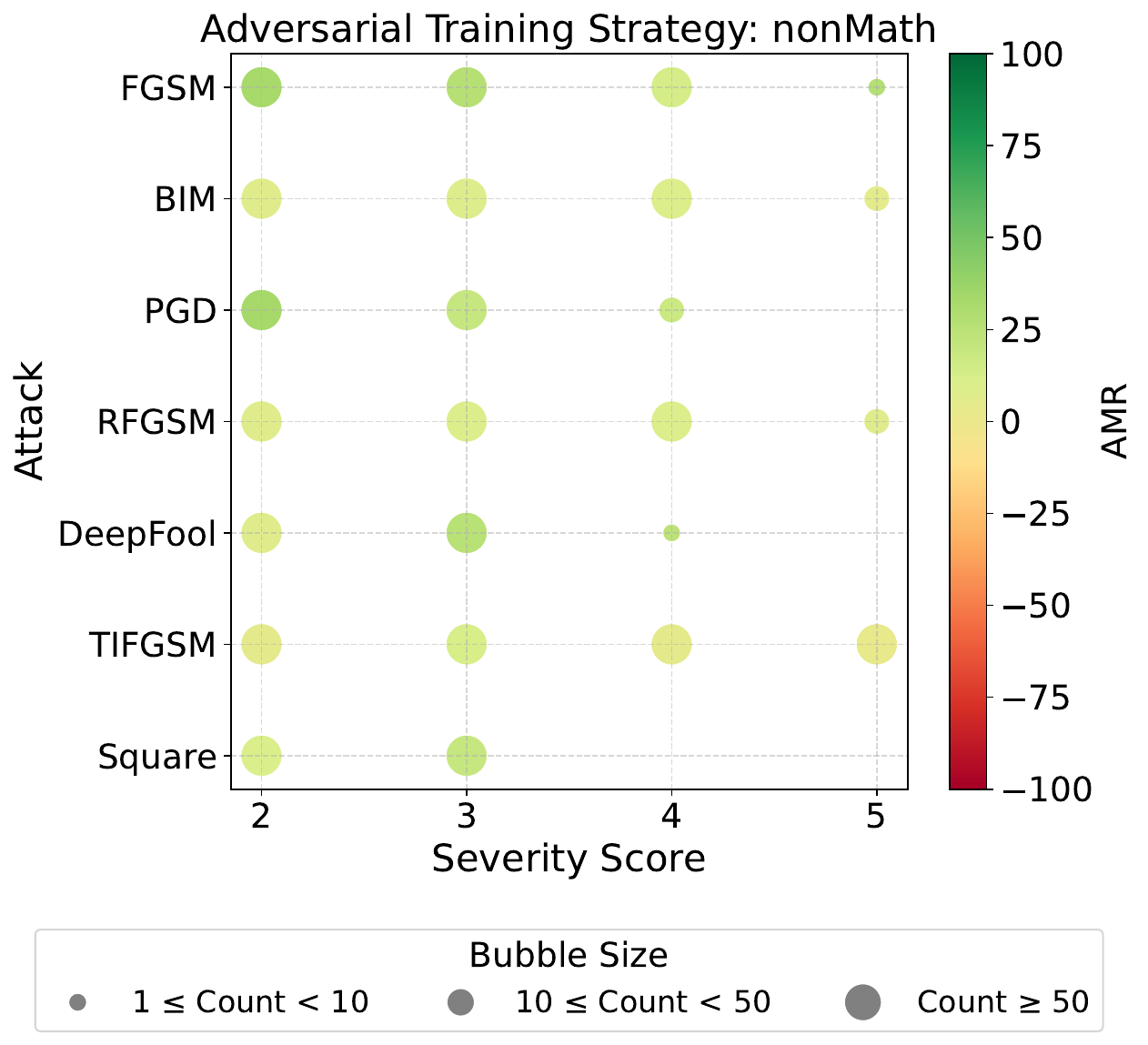}
        \caption{Attack severity score analysis.}
        \label{subfig:amr_nonmath_heatmaps_2x2_sev}
    \end{subfigure}
    \caption{\textit{Non-Math} training strategy AMR under different attacks.}
    \label{fig:amr_nonmath_heatmaps_2x2}
\end{figure}

\subsection{Adversarial Training Failiures}
\label{subsec:rq5}

% This section addresses Research Question 5 (RQ5): \textit{When and where does adversarial training fail? Specifically, how frequently does it result in a negative Attack Mitigation Rate (AMR), and are these failures uniformly distributed across scenarios C?} Understanding the conditions under which adversarial training degrades performance (i.e., results in a negative AMR) is crucial for practitioners deciding whether and how to implement these defenses. A negative AMR indicates that the model became \textit{more} vulnerable to a specific attack after undergoing adversarial training compared to the baseline model trained only on clean data.

% Overall, our extensive evaluation reveals that adversarial training is not always beneficial. Across all experiments, \textbf{20.53\% of evaluations resulted in a negative AMR}, indicating a non-trivial frequency of performance degradation. When these failures occur, the average negative AMR is \textbf{$-35.89\% (\pm 34.06$ standard deviation)}, suggesting that the drop in robustness can be substantial. Figure~\ref{fig:negativeAmrDistribution} illustrates the distribution of these negative AMR values, showing a concentration towards less extreme negative values but with a long tail representing significant degradations.

\begin{figure}
    \centering
    \includegraphics[width=0.625\linewidth]{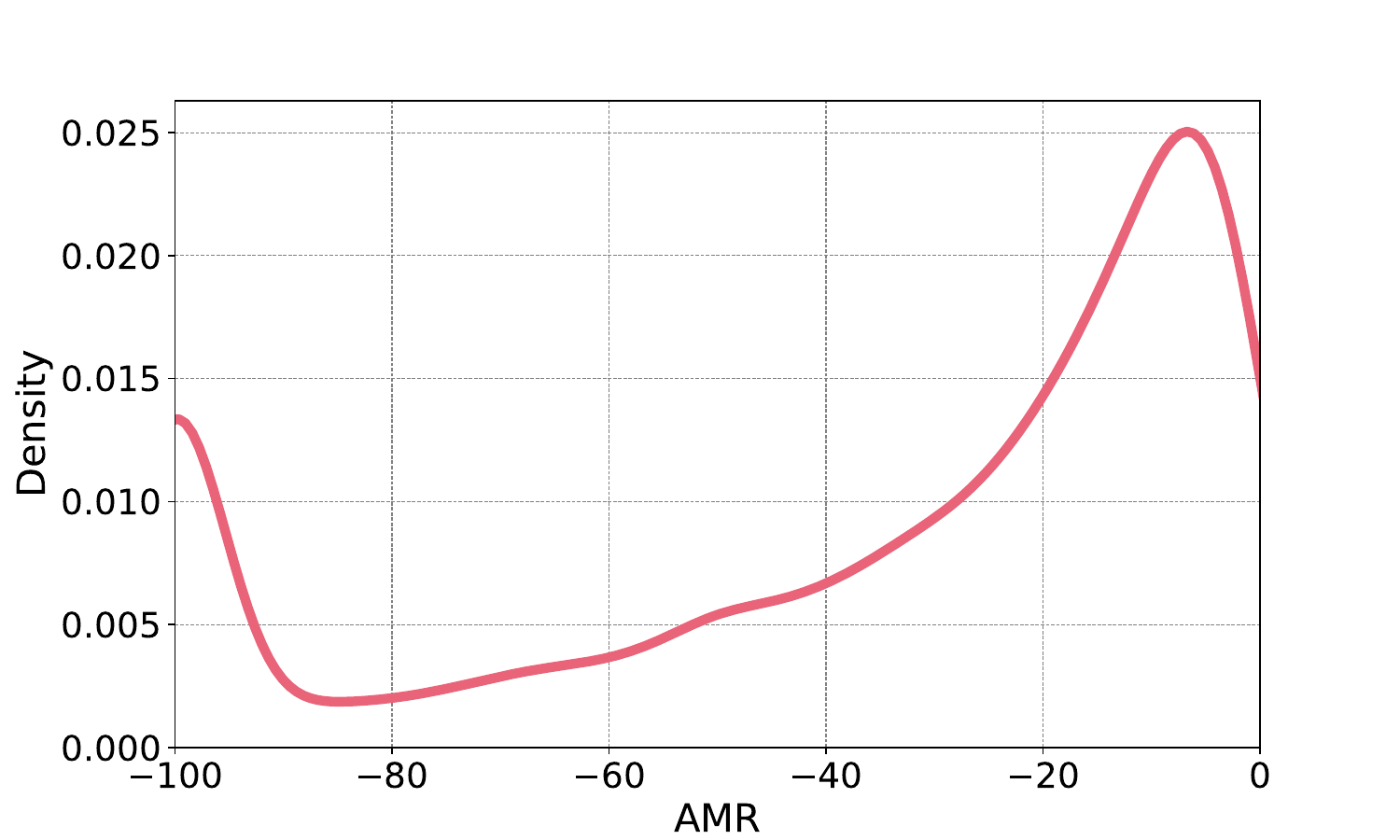}
    \caption{Negative AMR distribution}
    \label{fig:negativeAmrDistribution}
\end{figure}

Understanding when adversarial training harms rather than helps is critical for practitioners considering its adoption.
A negative AMR signals that the model has become more vulnerable to specific attacks than a baseline trained only on clean data.
Our evaluation shows that adversarial training is not universally beneficial: 20.53\% of all assessments resulted in a negative AMR, highlighting a notable risk of performance degradation.
When failures occur, the average negative AMR is $-35.89\% \pm 34.06$, indicating that the loss in robustness can be substantial.
As shown in Figure~\ref{fig:negativeAmrDistribution}, while most negative outcomes are moderate, practitioners must be aware of a long tail of severe degradations.

To better understand where adversarial training most often fails, we analyze the distribution of negative AMR across different experimental dimensions (Table~\ref{tab:failiures}).
Failures are heavily concentrated in DUMB scenarios involving a significant mismatch between attacker and victim models (Table~\ref{subtab:c}).
Notably, \texttt{C4} and \texttt{C8} (model and data mismatches) showed the highest rates of negative AMR, at 32.00\% and 28.70\%, respectively, while white-box settings (\texttt{C1}) exhibited minimal failure (1.61\%).
Different training strategies also varied in robustness (Table~\ref{subtab:advtr}): simple perturbation methods like \textit{Salt \& Pepper} and \textit{FGSM} led to frequent but less severe degradations, while the \textit{adaptive} strategy, despite strong overall performance, suffered the most severe failures when they did occur (-52.56\% on average).
Attack type also plays a key role (Table~\ref{subtab:atk}): \textit{DeepFool} and \textit{Square} attacks triggered failures most frequently, with \textit{DeepFool} causing the steepest average drop in robustness (-53.73\%).
Finally, failure rates were highly skewed toward weaker attacks (Table~\ref{subtab:sev}), with 72.53\% of negative AMR instances occurring against attacks initially classified as Severity 1; stronger attacks rarely caused adversarial training to backfire.
These results highlight that adversarial training, while powerful in some settings, is highly sensitive to mismatches, training strategies, attack types, and the initial strength of adversarial perturbations.

\begin{table*}[ht]
\centering
\caption{Values and distributions of negative AMR across experimental dimensions. ``\texttt{\% Neg.}'' indicates, for each dimension, the percentage of negative samples relative to the total number of negative samples across all dimensions., ``\texttt{Avg.}'' shows the mean AMR values and standard deviation, ``\texttt{Med.}'' indicates the median AMR.}
\resizebox{0.7\linewidth}{!}{%
\begin{tabular}{c}
\begin{minipage}{0.48\linewidth}
\centering
\subcaption{By DUMB case.}
\label{subtab:c}
\begin{tabular}{@{}l|ccc@{}}
\toprule
\textbf{Case} & \textbf{\% Neg.} & \textbf{Avg.} & \textbf{Med.} \\
\midrule
\textbf{C1} & 1.61\%  & $-35.45 \pm 35.85$ & $-20.00$ \\
\textbf{C2} & 7.45\%  & $-32.23 \pm 36.34$ & $-13.89$ \\
\textbf{C3} & 10.84\% & $-44.12 \pm 35.19$ & $-33.33$ \\
\textbf{C4} & 32.00\% & $-42.75 \pm 35.18$ & $-31.25$ \\
\textbf{C5} & 2.17\%  & $-32.88 \pm 35.10$ & $-15.79$ \\
\textbf{C6} & 7.44\%  & $-28.40 \pm 33.32$ & $-12.20$ \\
\textbf{C7} & 9.79\%  & $-29.97 \pm 30.01$ & $-18.18$ \\
\textbf{C8} & 28.70\% & $-30.28 \pm 30.87$ & $-17.24$ \\
\bottomrule
\end{tabular}
\end{minipage}
\hspace{0.04\linewidth}
\begin{minipage}{0.48\linewidth}
\centering
\subcaption{By adversarial training strategy.}
\label{subtab:advtr}
\begin{tabular}{@{}l|ccc@{}}
\toprule
\textbf{Train.} & \textbf{\% Neg.} & \textbf{Avg.} & \textbf{Med.} \\
\midrule
\textbf{FGSM}       & 14.70\% & $-34.40 \pm 33.41$ & $-20.69$ \\
\textbf{PGD}        & 9.11\%  & $-35.92 \pm 31.74$ & $-25.00$ \\
\textbf{Ada.}       & 12.61\% & $-52.56 \pm 36.15$ & $-45.83$ \\
\textbf{Cur.}       & 7.41\%  & $-43.09 \pm 34.35$ & $-32.58$ \\
\textbf{Ens.}       & 10.01\% & $-42.70 \pm 34.77$ & $-31.40$ \\
\textbf{Gau.}       & 11.81\% & $-27.55 \pm 31.87$ & $-13.04$ \\
\textbf{N.M.}       & 12.23\% & $-30.43 \pm 33.06$ & $-15.29$ \\
\textbf{S\&P}       & 14.01\% & $-21.80 \pm 25.82$ & $-11.54$ \\
\textbf{Sur.}       & 8.11\%  & $-42.31 \pm 34.69$ & $-30.71$ \\
\bottomrule
\end{tabular}
\end{minipage}
\\[0.5cm]
\begin{minipage}{0.48\linewidth}
\centering
\subcaption{By attack.}
\label{subtab:atk}
\begin{tabular}{@{}l|ccc@{}}
\toprule
\textbf{Attack} & \textbf{\% Neg.} & \textbf{Avg.} & \textbf{Med.} \\
\midrule
\textbf{BIM}      & 13.19\% & $-30.23 \pm 30.65$ & $-17.24$ \\
\textbf{DeepFool} & 24.30\% & $-53.73 \pm 38.02$ & $-47.37$ \\
\textbf{FGSM}     & 5.96\%  & $-22.38 \pm 24.88$ & $-12.31$ \\
\textbf{PGD}      & 9.41\%  & $-30.66 \pm 29.86$ & $-19.36$ \\
\textbf{RFGSM}    & 13.48\% & $-30.97 \pm 31.39$ & $-17.65$ \\
\textbf{Square}   & 22.39\% & $-40.41 \pm 32.41$ & $-30.36$ \\
\textbf{TIFGSM}   & 11.27\% & $-12.44 \pm 16.69$ & $-7.25$ \\
\bottomrule
\end{tabular}
\end{minipage}
\hspace{0.04\linewidth}
\begin{minipage}{0.48\linewidth}
\centering
\subcaption{By attack severity score.}
\label{subtab:sev}
\begin{tabular}{@{}l|ccc@{}}
\toprule
\textbf{Sev.} & \textbf{\% Neg.} & \textbf{Avg.} & \textbf{Med.} \\
\midrule
\textbf{1} & 72.53\% & $-43.13 \pm 35.13$ & $-30.77$ \\
\textbf{2} & 19.77\% & $-21.30 \pm 23.59$ & $-12.05$ \\
\textbf{3} & 4.77\%  & $-5.87 \pm 5.91$   & $-4.00$ \\
\textbf{4} & 1.50\%  & $-4.93 \pm 4.78$   & $-3.29$ \\
\textbf{5} & 1.43\%  & $-2.72 \pm 3.08$   & $-1.29$ \\
\bottomrule
\end{tabular}
\end{minipage}
\end{tabular}
}
\label{tab:failiures}
\end{table*}

\section{Conclusions}
\label{sec:conclusions}

This paper examined the effectiveness of various adversarial training strategies across the scenarios defined by the DUMB framework.
Based on 130k evaluations across all folds of our updated and adapted attacker model, we provide the following answers to our research questions.
% \begin{enumerate}[label=\textbf{A\arabic*:}, leftmargin=*, align=left]
\begin{enumerate}\setlength{\itemindent}{1em}
    \item[\textbf{A1:}] Attack strategies designed for transferability (e.g., TIFGSM) pose a greater threat to the overall DUMB population, though non-mathematical, model-agnostic methods also demonstrate significant ASRs.
    \item[\textbf{A2:}] When the threat model is unknown, strategies such as \textit{adaptive} and \textit{curriculum} generally yield the highest AMRs. However, adversarial training can slightly degrade performance on lower-impact attacks in sub-optimal tasks.
    \item[\textbf{A3:}] Adversarial training proves most effective when the source and target models align. This also holds for mismatched datasets, but only when the original task is sub-optimal in baseline evaluation. Performance can degrade in scenarios with multiple mismatches, particularly when attacks have lower ASRs.
    \item[\textbf{A4:}] The most effective adversarial training strategies generalize across different target attacks, though their success varies depending on the attacker's knowledge. However, specific attacks (e.g., \textit{DeepFool} and \textit{Square}) can negatively impact adversarially trained models, highlighting the need for focused attention on these cases.
    \item[\textbf{A5:}] Adversarial training is less effective against lower severity attacks, indicating that it should be employed primarily when facing significantly impactful threats.
\end{enumerate}

\paragraph{Future Works.}
% Future research could explore more sophisticated adversarial training strategies that address the limitations observed in lower severity attacks.
% Additionally, investigating hybrid approaches combining model-agnostic techniques with transferability-optimized strategies may offer more robust defenses across diverse attack scenarios.
% Expanding the evaluation to include additional threat models and real-world datasets would provide further insight into the generalizability of these defenses.
% Lastly, a deeper analysis of the trade-offs between attack severity and training strategies could help fine-tune adversarial defenses for specific practical applications.
Future research could develop advanced adversarial training strategies to overcome weaknesses against low-severity attacks.
Exploring hybrid defenses that combine model-agnostic and transfer-optimized methods may further enhance robustness across diverse scenarios.
Expanding evaluations to broader threat models and real-world datasets would improve understanding of defense generalizability.
Finally, a deeper study of the trade-offs between attack severity and training strategies could guide the design of defenses tailored to specific applications.

% ---- Bibliography ----
\bibliographystyle{splncs04}
\bibliography{bibliography}

\appendix
\section{Additional Results}
\label{app:results}

We now provide more details on the results shown in Section~\ref{app:results}.

\subsection{RQ1}
\label{app:rq1}

\begin{figure}[htbp]
    \centering
    \begin{subfigure}{\linewidth}
        \centering
        \includegraphics[width=\linewidth]{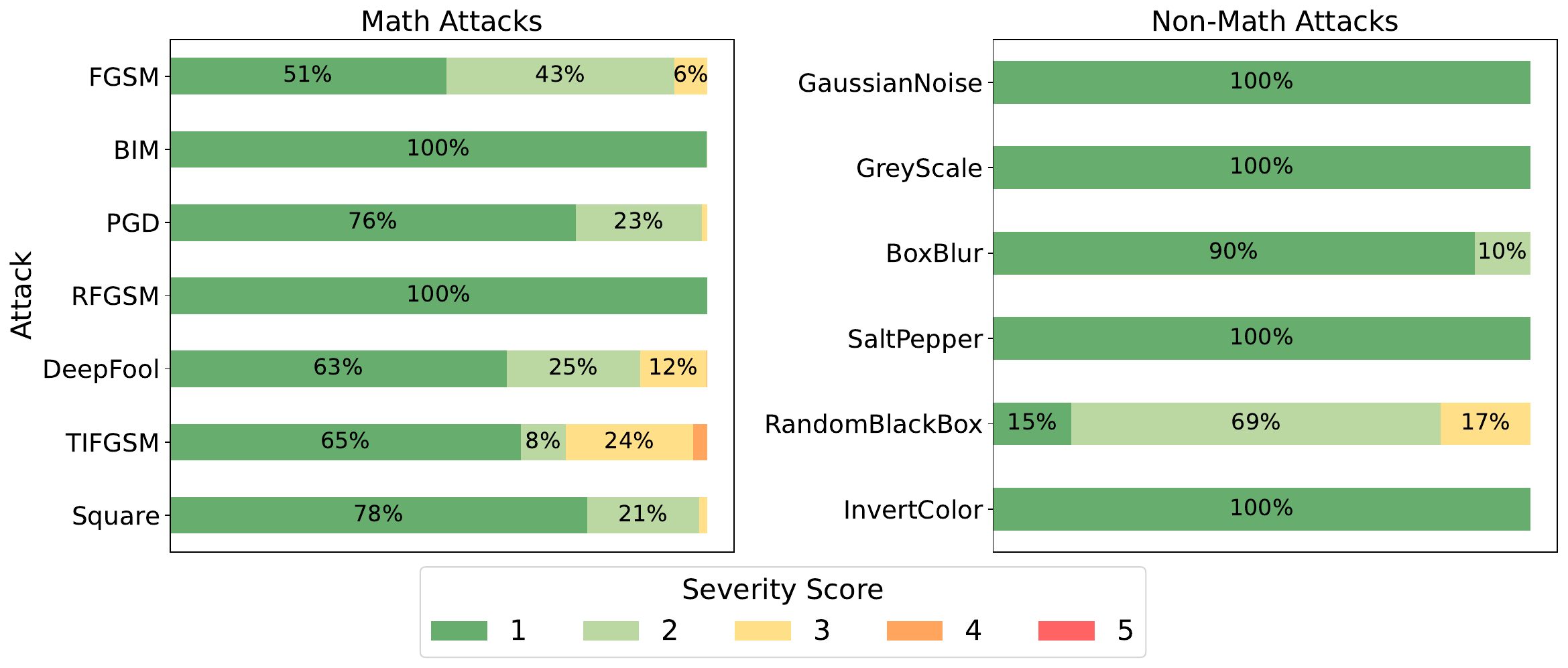}
        \caption{B\&M.}
        \label{subfig:rq2task1}
    \end{subfigure}
    \begin{subfigure}{\linewidth}
        \centering
        \includegraphics[width=\linewidth]{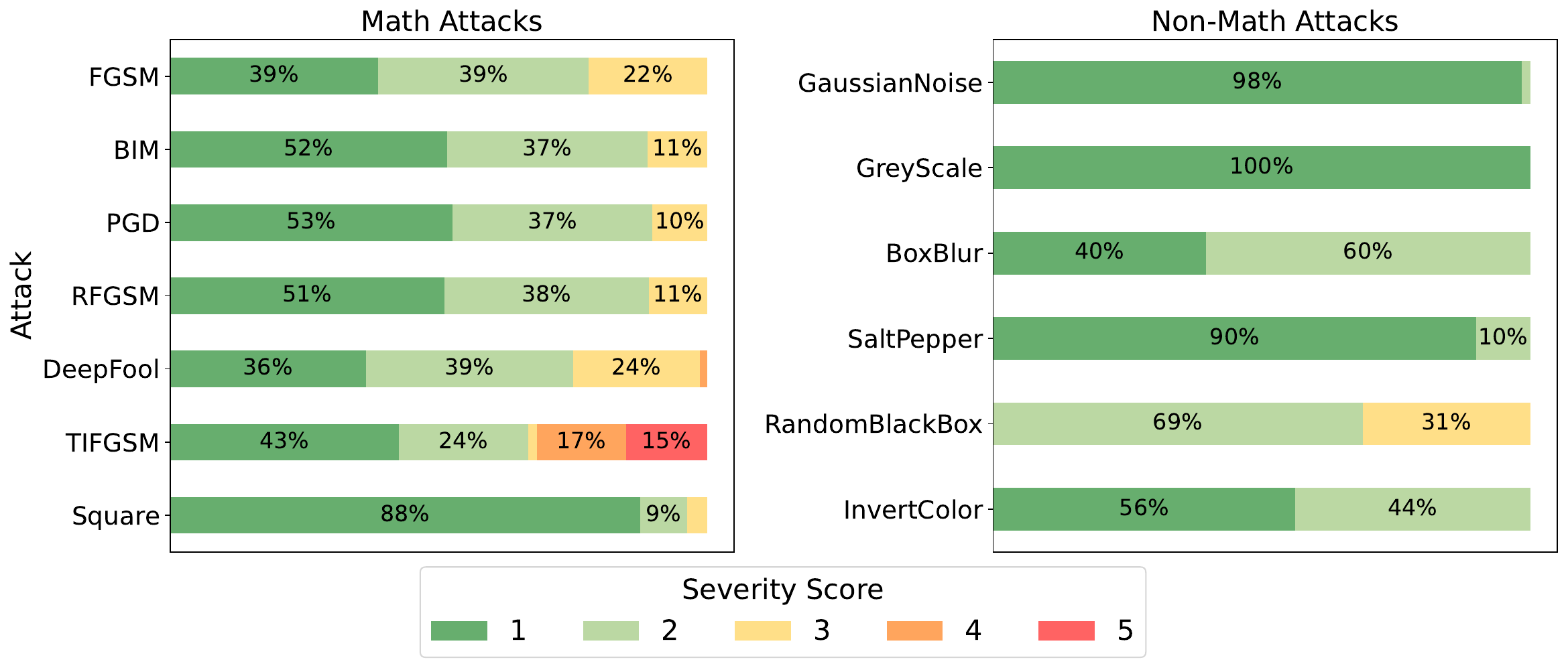}
        \caption{C\&D.}
        \label{subfig:rq2task2}
    \end{subfigure}
    \begin{subfigure}{\linewidth}
        \centering
        \includegraphics[width=\linewidth]{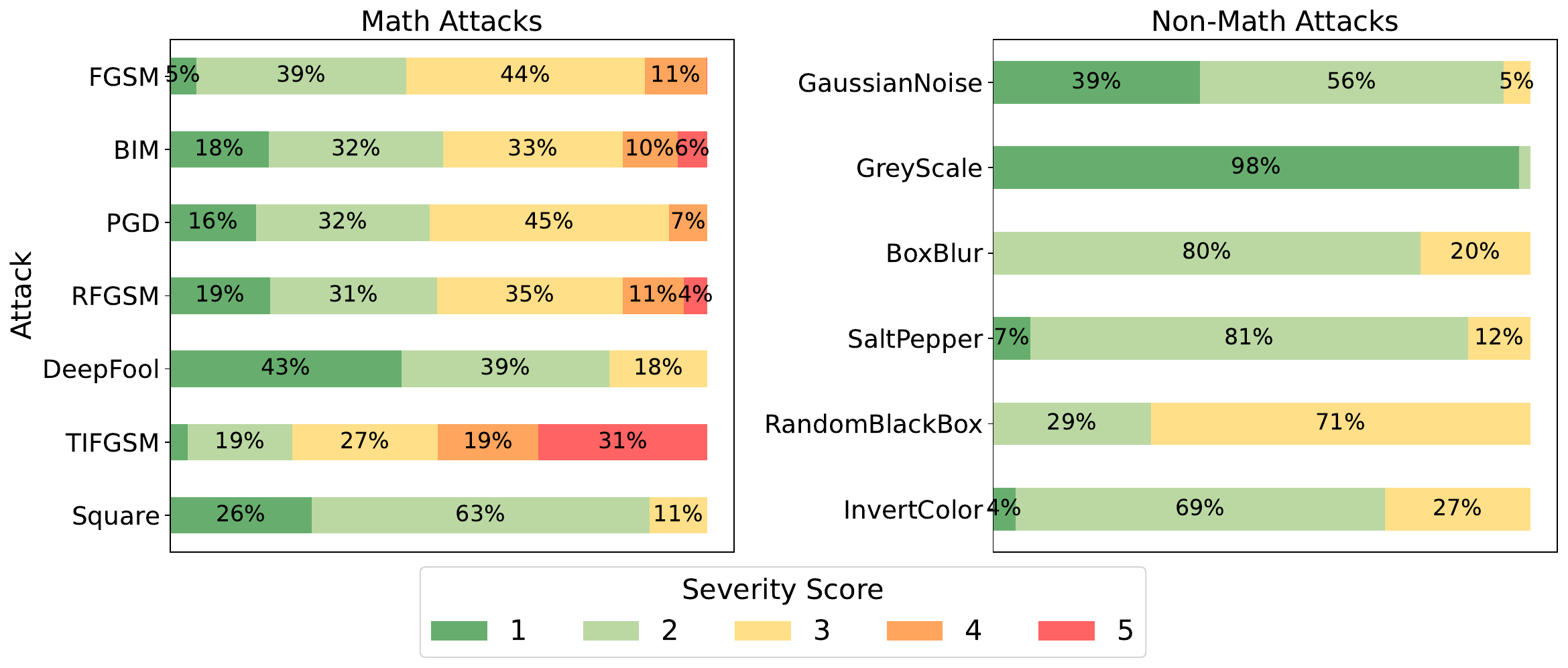}
        \caption{M\&W.}
        \label{subfig:rq2task3}
    \end{subfigure}
    \caption{Severity score distribution of each attack on each task.}
    \label{fig:rq1tasks}
\end{figure}

While Section~\ref{subsec:rq1} provides a cross-task overview of attack severity scores, Fig.~\ref{fig:rq1tasks} breaks them down by individual task.
As noted in the original DUMB paper, attacks tend to become more effective as tasks grow more challenging.
Specifically, the more straightforward task (B\&M) shows no high-severity mathematical attacks, whereas the most complex task (M\&W) features many.
A similar trend appears for non-mathematical attacks, although their overall ASR remains generally lower.

% \subsection{RQ2}
% \label{app:rq2}

% \subsection{RQ3}
% \label{app:rq3}

\subsection{RQ4}
\label{app:rq4}

\begin{figure}[htbp]
    \centering
    \begin{subfigure}{0.475\linewidth}
        \centering
        \includegraphics[width=\linewidth]{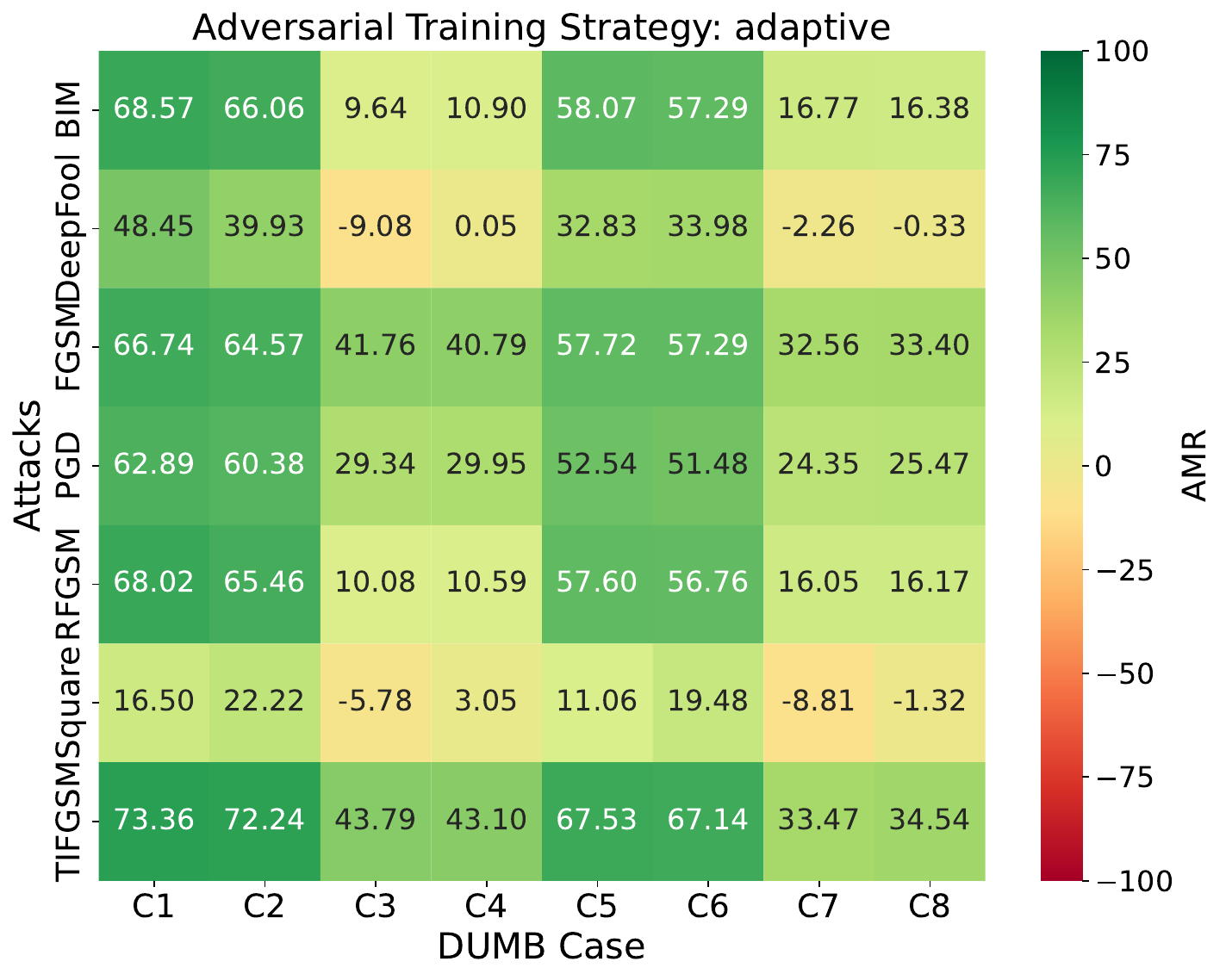}
        \caption{DUMB cases analysis.}
        \label{subfig:amr_ada_heatmaps_2x2_dumbnon}
    \end{subfigure}
    \hfill
    \begin{subfigure}{0.475\linewidth}
        \centering
        \includegraphics[width=\linewidth]{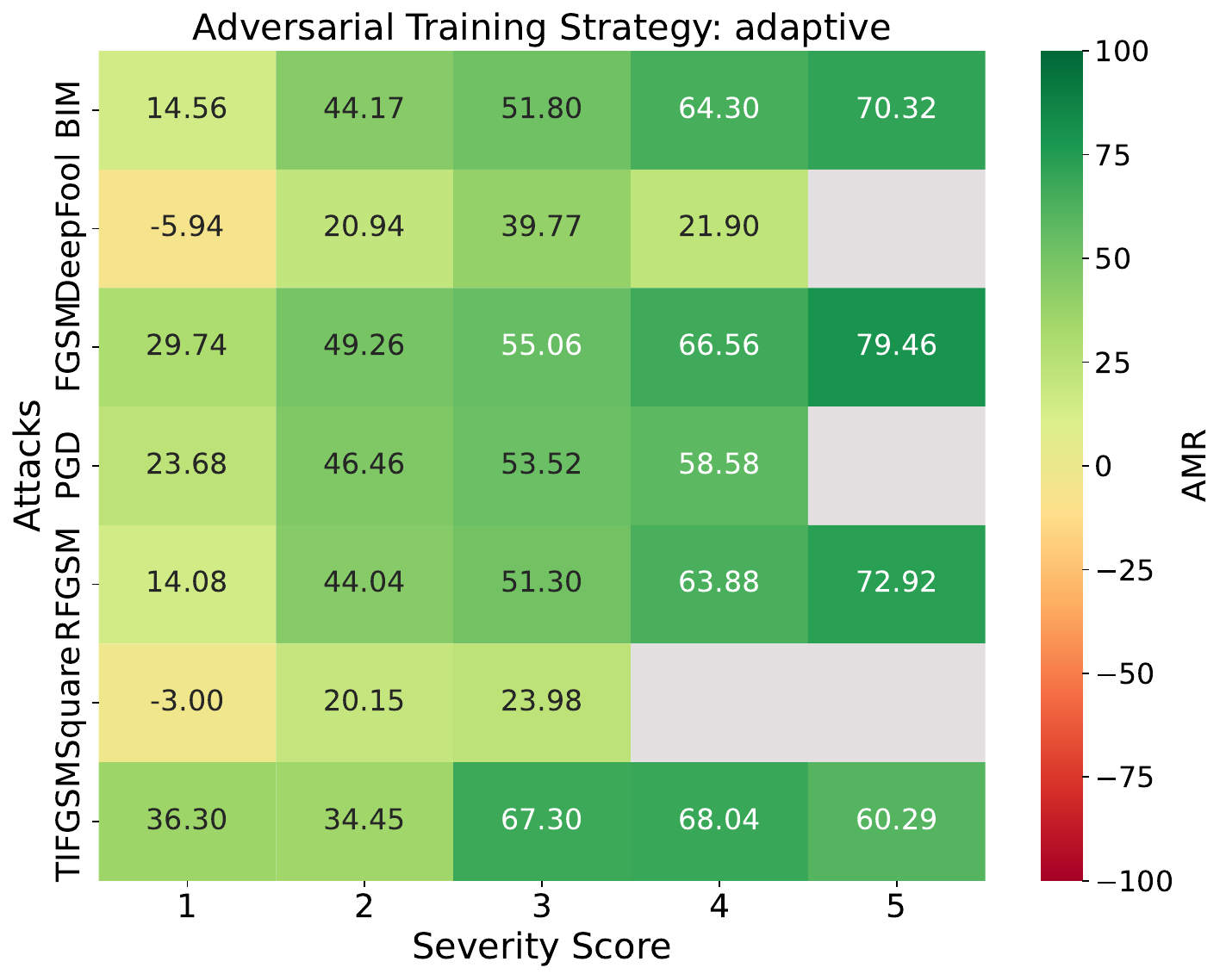}
        \caption{Attack severity score analysis.}
        \label{subfig:amr_ada_heatmaps_2x2_sevnon}
    \end{subfigure}
    % \vskip\baselineskip
    \caption{\textit{Adaptive} training strategy AMR under different attacks.}
    \label{fig:amr_ada_heatmaps_2x2non}
\end{figure}

\begin{figure}[htbp]
    \centering
    \begin{subfigure}{0.475\linewidth}
        \centering
        \includegraphics[width=\linewidth]{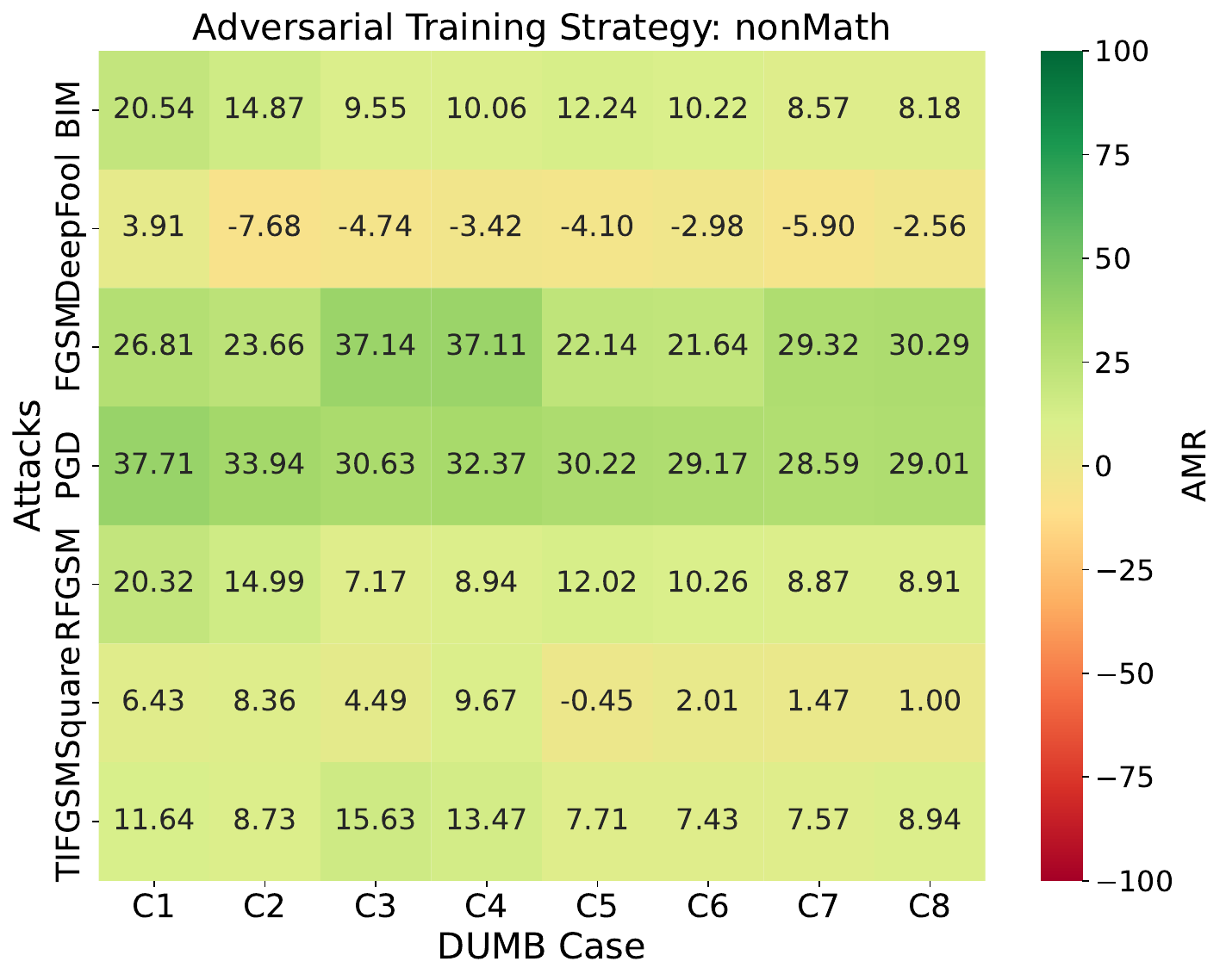}
        \caption{DUMB cases analysis.}
        \label{subfig:amr_nonmath_heatmaps_2x2_dumbnon}
    \end{subfigure}
    \hfill
    \begin{subfigure}{0.475\linewidth}
        \centering
        \includegraphics[width=\linewidth]{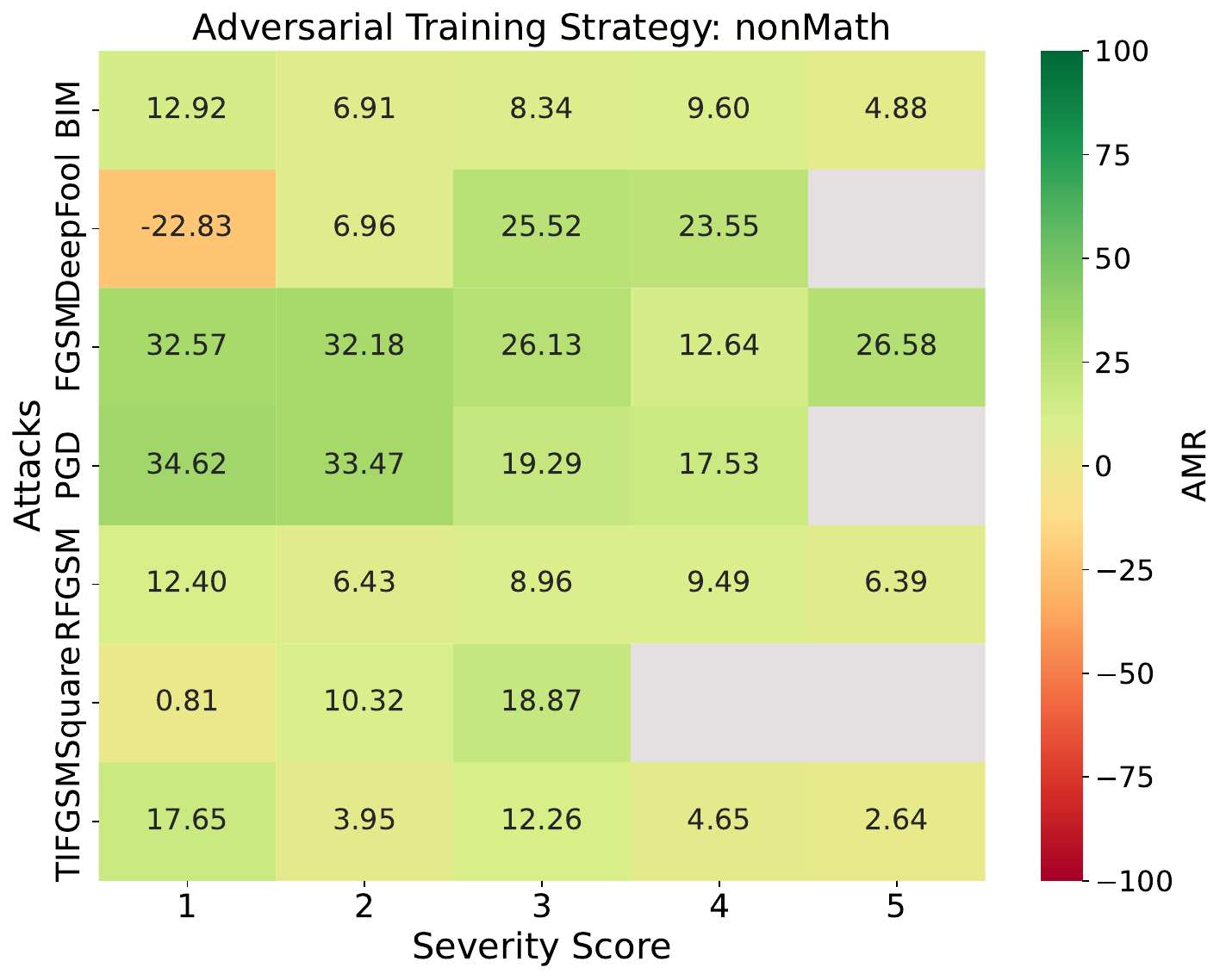}
        \caption{Attack severity score analysis.}
        \label{subfig:amr_nonmath_heatmaps_2x2_sevnon}
    \end{subfigure}
    \caption{\textit{Non-Math} training strategy AMR under different attacks.}
    \label{fig:amr_nonmath_heatmaps_2x2non}
\end{figure}

In Fig.~\ref{fig:amr_ada_heatmaps_2x2non} and Fig.~\ref{fig:amr_nonmath_heatmaps_2x2non}, we extend the analysis from Section~\ref{subsec:rq4} by including the exact AMR values for each case.

% \vspace{-15mm}

% \subsection{RQ5}
% \label{app:rq5}

\end{document}